\documentclass[]{aa}  

\usepackage{xcolor}

\usepackage{graphicx}
\usepackage{txfonts}
\usepackage{lipsum}
\usepackage{subcaption}         
\usepackage{lscape}             
\usepackage{placeins}           
                                
\usepackage[colorlinks=true, linkcolor=blue, citecolor=blue, urlcolor=blue]{hyperref}
\usepackage{url}
\usepackage{soul}

\newcommand{\micron}{$\mu$m}

\def\kms{{\text{km\,s}$^{-1}$}}

\def\Rsun{{\rm R$_{\odot}$}}
\def\Msun{{\rm M$_{\odot}$}}

\begin{document}

\title{Late-Time Evolution of the Massive Stellar Merger M101 OT2015-1}


%
\authorrunning{G{\'o}mez-Mu{\~n}oz et al.}
\author{Marco~A. G{\'o}mez-Mu{\~n}oz \inst{1,2,3} \corrauth{mgomez@icc.ub.edu}
\and Nadejda Blagorodnova \inst{1,2,3} \email{nblago@fqa.ub.edu}
\and Tomasz Kami{\'nski} \inst{4} \email{tomkam@ncac.torun.pl}
\and Gerard Garcia-Moreno \inst{1,2} \email{ggarcimo@fqa.ub.edu}
\and Maxime Wavasseur \inst{1,2} \email{m.wavasseur@icc.ub.edu}
\and Hugo Tranin\inst{1,2} \email{htranin@icc.ub.edu}
\and Grace Katusiime \inst{1,2,3} \email{gracekatusiime@icc.ub.edu}
\and Jacob~E. Jencson \inst{5} \email{jjencson@ipac.caltech.edu}
\and Mansi~M. Kasliwal \inst{6} \email{mansi@astro.caltech.edu}
}

\institute{Departament de F{\'i}sica Qu{\`a}ntica i Astrof{\'i}sica (FQA),       Universitat de Barcelona (UB),  c. Mart{\'i} i Franqu{\`e}s, 1, 08028 Barcelona, Spain
    \and Institut de Ci{\`e}ncies del Cosmos (ICCUB), Universitat de Barcelona (UB), c. Mart{\'i} i Franqu{\`e}s, 1, 08028 Barcelona, Spain
    \and Institut d'Estudis Espacials de Catalunya (IEEC), Edifici RDIT, Campus UPC, 08860 Castelldefels (Barcelona), Spain
    \and Nicolaus Copernicus Astronomical Center, Polish Academy of Sciences, ul. Rabia{\'n}ska 8, 87-100 Toru{\'n}, Poland
    \and IPAC, Mail Code 100-22, Caltech, 1200 East California Boulevard, Pasadena, CA 91125, USA
    \and Division of Physics, Mathematics and Astronomy, California Institute of Technology, Pasadena, CA 91125, USA
}

\date{Received September 30, 20XX}

\abstract{Luminous red novae (LRNe) are astronomical transients arising from unstable mass transfer and common envelope phases in close binary systems, eventually culminating in stellar mergers. Their cold, expanding ejecta provide the ideal conditions for molecule and dust formation. We investigate the late-time dust and molecular evolution of the LRN M101\,OT2015-1 (M101-OT), a high-mass LRN progenitors ($18 \pm 1\,\text{M}_\odot$), using multi-band near-infrared (NIR) and mid-infrared (MIR) observations spanning up to $\sim 5$ years post-outburst.
Archival photometry from NEOWISE, Spitzer, Keck/NIRC2, and Keck/MOSFIRE are interpolated via two-dimensional Gaussian Process regression. Spectral energy distributions (SEDs) are modelled with 1D spherical radiative transfer models to track the evolution of dust mass, dust temperature, and central source properties. At day $+150$ post-peak, the SED is well-fit by a pure stellar blackbody ($T_{\rm eff} \approx 3330\text{ K}$) with negligible dust ($\log M_{\rm dust}/\text{M}_\odot < -6.36$). Rapid dust condensation begins around day $+200$ ($T_{\rm dust} \approx 1591\text{ K}$), with dust mass increasing to $\log M_{\rm dust}/\text{M}_\odot \approx -3.65$ and optical depth reaching $\tau_V \approx 64$ by day $+1300$. A prominent NIR re-brightening at day $\sim 600$ ($M_K \approx -12.04$), accompanied by a dust reheating bump ($T_{\rm dust} \approx 1567\text{ K}$), highlights a second phase of dust nucleation likely driven by shock interactions.
Low-resolution NIR MOSFIRE spectra are analysed for line identification and also using an LTE isothermal slab model to characterise CO molecular features. The MOSFIRE spectra reveal an O-rich environment with photospheric expansion velocities of $\sim520\text{--}620\text{ km s}^{-1}$, as measured in the atomic \ion{Fe}{i}, \ion{He}{i}, Pa$\beta$ emission lines, and high CO column densities ($\log N_{\rm CO}/\text{cm}^{-2} = 20.9$ and $20.7$ at $+142\text{ d}$ and $+197\text{ d}$, respectively). These results confirm that LRNe are prolific producers of cosmic dust and are efficient molecular factories.
}


\keywords{Stellar mergers --
Astrophysical dust process --
Infrared photometry --
Infrared spectroscopy --
Circumstellar dust --
stars:individual:M101\,OT2015-1
}

\maketitle
\nolinenumbers

\section{Introduction}

The majority of the intermediate- and high-mass stars ($\sim$80\%) in the Galaxy
are members of a binary or multiple stellar system \citep{2012Sci...337..444S,2017ApJS..230...15M}.
A significant fraction of massive stars ($\sim$70\%) will have an interaction
with its companion.
These interactions often culminate
in a common envelope (CE) evolution phase, leading to a compact binary, or a stellar merger
\citep{1976IAUS...73...75P,2017ApJ...843L..30I}.
While theoretically predicted for decades, the observational signature of these
catastrophic events has been identified under the class of luminous
red novae (LRNe). These astrophysical transients serve as the primary laboratory for studying
non-conservative mass transfer that dictates the final fate of binary systems
\citep[e.g.][and references therein]{2014ApJ...788...22P,2025ApJ...994L..41K};
from the formation of exotic compact objects
\citep[such as gravitational wave sources; e.g.][]{2000A&A...360.1011N,2017ApJ...846..170T,2018A&A...615A..91B,2018MNRAS.481.4009V}
to the enrichment of the galactic
environment \citep[e.g.][]{2020MNRAS.496.5503B,2026ApJ...999L..35B,2026A&A...706A.154R,2026ApJ...999...16K}.

Luminous red novae are characterised by a distinct multi-peaked light curve, with
luminosities between novae and supernovae (10$^{38}$ to 10$^{41}$\,erg\,s$^{-1}$),
and the expansion
of the progenitor's envelope, which leads to a dramatic cooling of the photosphere
transitioning from optical to infrared (IR) dominance \citep[e.g.][]{2017ApJ...834..107B,2020MNRAS.496.5503B,2022A&A...667A...4C,2023A&A...671A.158P,KaminskiBlagorodnova2026}. 
The low-velocity and cold ejecta from these violent binary interactions are ideal sites for dust formation, as shown by studies at IR and mm wavelengths
\citep{2018A&A...617A.129K,2020MNRAS.496.5503B,2025A&A...699A.316S,2026ApJ...999...16K,2026A&A...706A.154R}.  

LRNe coming from massive systems have
volumetric rates comparable to core-collapse supernovae
\citep[CCSNe; $\sim$77\%,][]{2023ApJ...948..137K} and are expected to eject a significant fraction of the primary star's envelope mass \citep[e.g.][]{2022ApJ...938....5M}. 
Simulations have
predicted that the most frequent, lower mass systems will contribute with dust
masses comparable to those delivered by asymptotic giant branch (AGB) stars
\citep{2024MNRAS.527.9145G,2024MNRAS.533..464B}.  
While dust condensation in LRNe occurs on timescales of months to years, broadly comparable to those observed in SN ejecta, AGB stars produce dust trough continuous mass loss over much longer evolutionary timescales ($\sim10^{4}-10^{6}$\,yr).

AGB stars and CCSNe have been traditionally viewed as the primary sources of galactic dust.
However, the efficiency of CCSNe as a dust producer remains debated because of the destructive nature of the reverse shock in which dust grains are destroyed \citep{2016A&A...590A..65M,Kirchschlager2022}.
Recent studies have shown that LRNe could represent a similarly important dust production due to
the environment in which they take place. 
For example, late-time characterization of a representative LRN sample—leveraging the sensitivity of the JWST
Mid-Infrared Instrument (MIRI) and Low-Resolution Spectrograph (LRS)—has revealed that these mergers are prolific dust producers.
Specifically, the analysis of AT\,2021blu, AT\,2021biy, AT\,2018bwo, and M31-LRN-2015 has yielded dust masses
of 4.2$\times$10$^{-5}$, 30$\times$10$^{-5}$, 7.5$\times$10$^{-5}$, and 77$\times$10$^{-5}$ M$_\sun$, respectively.
While the less-massive, Galactic, red novae are generally thought to disperse less dust, the remnant of BLG360 has produced 1.2$\times$10$^{-3}$ M$_\sun$ of dust \citep{2025A&A...699A.316S}.
These values suggest that LRNe may contribute up to 25\% as much dust as CCSNe to the local cosmic dust budget
\citep{2026ApJ...999...16K}. Although the dust production may occur on shorter timescales, this could help ease
the “dust budget crisis”—the apparent excess of ISM dust in our Galaxy relative to the expected contributions
from AGB stars and CCSNe \citep{Ginolfi2018}—and may be relevant to the early Universe as well \citep[e.g. to account for massive star-forming galaxies and their contribution to the start-formation history;][]{2026ApJ...998L..36Z}.

LRNe also serve as prolific
laboratories for complex molecule formation. High-resolution spectroscopy has revealed 
rich gaseous environments containing species such as AlO, VO, SiO, CO, and water \citep[e.g.][]{2004ApJ...615L..53B,2026ApJ...999L..35B}, all 
of which are characteristic signatures of oxygen-rich (O-rich) mineralogy \citep[e.g.][]{2015ApJ...814..109B}. 
These molecules are critical as they act as the primary building blocks for dust; for instance, 
the detection of Si-rich dust at $\sim$2 years post-outburst in Galactic LRNe--such as V4332 Sgr 
and V1309 Sco--and the detection of alumina dust \citep{2007ApJ...666L..25B,2021AJ....162..183W}, marks the efficient transition from the gas phase to solid-state grains in the cooling merger ejecta.

Despite the clear importance of LRNe as a cosmic dust factories, systematic long-term monitoring in the near-IR (NIR) and mid-IR (MIR) is scarce. 
This is particularly important for extragalactic transients originating from massive progenitors which go
through different thermodynamic envelope evolutions \citep{2026ApJ...999...16K} compared to their lower-mass counterparts
\citep[such as the well studied M31-LRN-2015;][]{2020MNRAS.496.5503B,2026ApJ...999...16K}.
Furthermore, late-time spectroscopic data for these distant objects are rare due to rapid optical fading.
Although NIR spectra exist for a few extragalactic massive-progenitor LRNe at different post-outburst epochs \citep[e.g.][]{2023ApJ...948..137K,2021A&A...653A.134B,Reguitti2026}, no attempts have been made to quantitatively
model these late-time gaseous environments. 
Consequently, it remains poorly understood what physical conditions prevail within the expanding outer layer, and how key dust-forming chemical building blocks such as CO survive the outburst.

To bridge this gap, this paper presents a comprehensive $\sim$5\,year photometric and 
spectroscopic analysis of the post-outburst evolution of the massive stellar merger LRN M101\,OT2015-1.
The LRN M101\,OT2015-1 (hereafter M101-OT, also known as M101-2015OT1, AT\,2015dl or PSN J14021678+5426205), first discovered by C. D. V{\`i}ntdevara at the B{\^a}rlad Astronomical Observatory on 10 February 2015\footnote{\url{http://www.cbat.eps.harvard.edu/unconf/followups/J14021678+5426205.html}} and published by 
\citet{Goranskij2016AstBu}, is located at the outskirts of the spiral galaxy M101 ($\alpha = 14^{\text{h}}02^{\text{m}}16.78^{\text{s}}$, $\delta = +54^{\circ}26'20.5''$), in the arm situated $3'41''$ north and $8'12''$ west from the nucleus.
Initial imaging and subsequent follow-up campaigns confirmed its nature as a remarkably
massive member of the LRNe class \citep{2017ApJ...834..107B,Pastorello2019}.
Detailed analysis of archival pre-outburst data revealed a long-lasting $\sim$5\,year precursor
brightening phase, whose properties resembled that of an F-type yellow
supergiant progenitor with an estimated mass of 18$\pm$1\,{\Msun} \citep{2017ApJ...834..107B}.
The primary physical and observational parameters of M101-OT are summarized in Table~\ref{tab:m101_properties}.
While the MIR light curve evolution of the massive merger M101\,OT
has been recently studied by \citet{2026A&A...706A.154R}, a detailed modelling of their dust evolution over
extended baselines has not been performed.

In addition, the early post-outburst optical light curve of M101-OT exhibited a double-peaked profile,
which places this object into the `risers' group of LRNe \citep{KaminskiBlagorodnova2026}.
This group represents systems whose observed light curves are interpreted as the result of the
merger ejecta interacting with dense, pre-existing circumstellar material, rather than being
dominated by recombination-powered plateaus as seen in lower-mass stellar mergers
\citep{2017MNRAS.471.3200M,2025ApJ...994L..41K}.

Here, we present a photometric and spectroscopic
analysis of the late-time evolution of M101-OT by using archival and unpublished
NIR and MIR data.
This photometric follow-up shows a re-brightening in the NIR bands
uncovering a complex envelope evolution likely powered by circumstellar shock interactions.
We also present a multi-epoch spectral energy distribution analysis by means of a radiative
transfer modelling, tracing the evolution of the dust parameters, such as dust mass, dust temperature,
and the overall bolometric luminosity.
Finally, we present the first quantitative, local thermodynamic equilibrium (LTE) isothermal slab modelling
of the late-time CO first-overtone molecular absorption of M101-OT.
Through this modelling, we show that near-complete CO molecular saturation is achieved just a few hundred
days post-outburst, providing unique insights into the rapid chemical pathways, molecular survival,
and density dilution of expanding merger ejecta.

\begin{table}
\centering
\caption{Physical and observational properties of M101~OT2015-1. \label{tab:m101_properties}}
\begin{tabular}{lcc}
\hline\hline
Parameter & Value & Reference \\
\hline
Right Ascension (J2000)      & $14^{\rm h}02^{\rm m}16.78^{\rm s}$                  & (1) \\
Declination (J2000)          & $+54^\circ26'20.5''$                                  & (1) \\
Distance ($D$)               & $6.4 \pm 0.2$\,Mpc                                     & (2) \\
$E(B-V)$ & $0.008$ mag                                           & (3) \\
Progenitor Classification    & F-type                            & (4) \\
Progenitor Mass              & $18${\Msun}                                & (4) \\
Precursor Duration           & $\sim 5$ years                                        & (4) \\
Light Curve Class            & Riser                             & (5) \\
Redshift ($z$) & 0.000804 & (6)  \\
\hline
\end{tabular}
\tablebib{
(1) \citet{Goranskij2016AstBu}; 
(2) \citet{2011ApJ...733..124S}; 
(3) \citet{2011ApJ...737..103S};
(4) \citet{2017ApJ...834..107B};
(5) \citet{KaminskiBlagorodnova2026};
(6) \citet{2023ApJ...955L...8H}
}
\end{table}

\begin{figure*}
    \centering
    \includegraphics[width=0.8\linewidth]{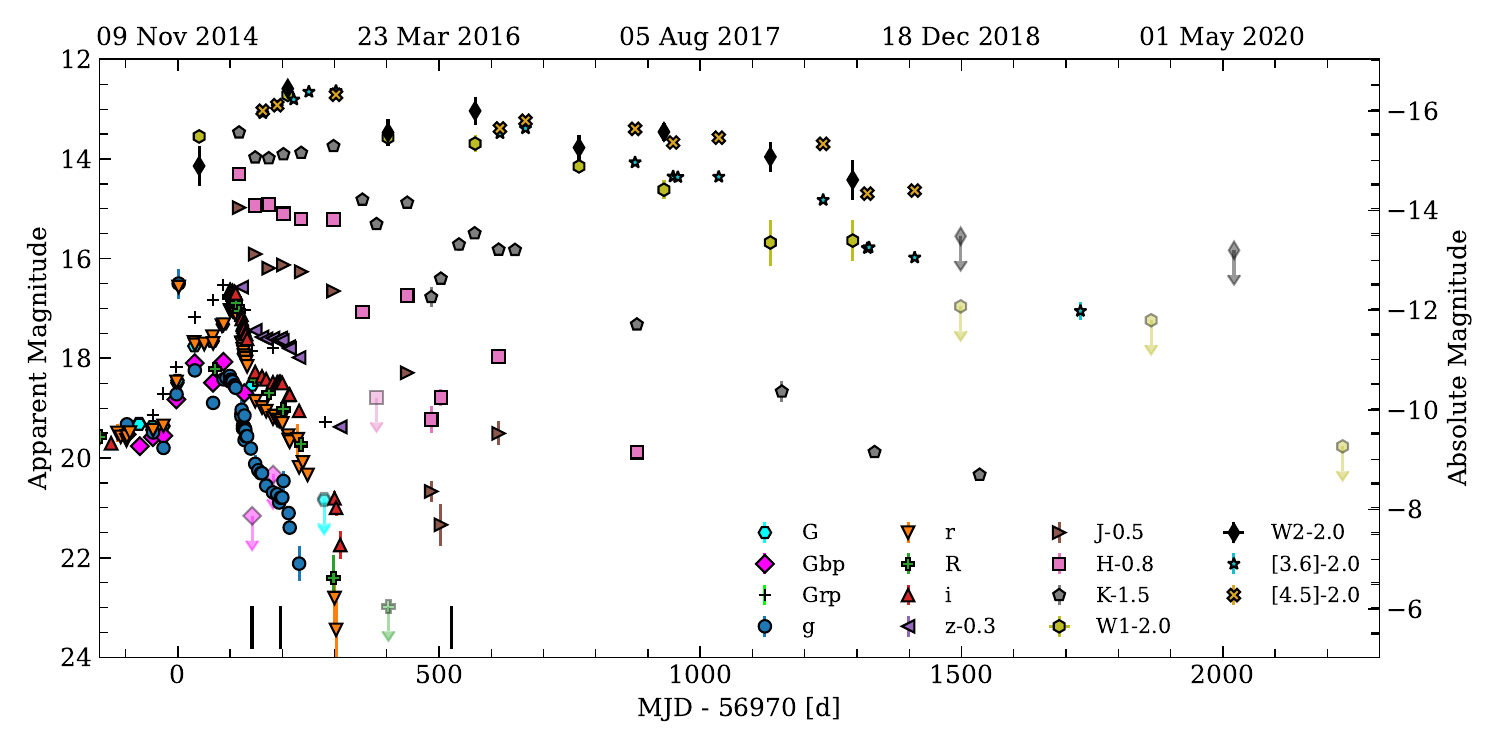}
    \caption{Multi-band light curve of the LRN M101-OT spanning $\sim$4.8\,yr after the first peak (56970 MJD in the $r$-band). Optical \textit{griz} bands are in AB magnitude system while the $R$, Gaia, and IR bands are in the Vega system. MJD errorbars are set for NEOWISE co-adds. The vertical solid lines indicate the date of the MOSFIRE NIR spectra. Upper-limits are shown as a pointing down arrows. \label{fig:figure1} 
    }
\end{figure*}


\section{Observational data}
\label{sec:obs_data}

To improve the analysis of the late-time evolution of M101-OT, we downloaded the existing  
NEOWISE and Spitzer photometry and include previously unpublished data from NIR 
Camera 2 (NIRC2) and the MOSFIRE photometric camera both mounted at the Keck II telescope.
Additionally, we also downloaded and processed unpublished NIR 
spectra taken with the MOSFIRE instrument.

\subsection{NEOWISE data}

NEOWISE \citep{2011ApJ...731...53M,2014ApJ...792...30M} has imaged the field of M101-OT for almost 14 years in both \textit{W1} ($\lambda_\mathrm{eff}=3.35$\,{\micron}) and \textit{W2} ($\lambda_\mathrm{eff}=4.60$\,{\micron}) bands; from 8 June 2010 to 24 May 2024. We downloaded all the single exposure (Level 1b) \textit{W1} and \textit{W2} images from the NASA/IPAC Infrared Science Archive (IRSA\footnote{\url{https://irsa.ipac.caltech.edu/frontpage/}}).
Given the sparse cadence of the observations, we grouped the data per season and co-add them using the {\sc MontagePy} package (a python binary extensions of existing {\sc montage} modules\footnote{{\sc montage}  is a toolkit for assembling Flexible Image Transport System (FITS) images into custom mosaics (see \url{http://montage.ipac.caltech.edu/}).}).
We subtracted the overall background of each co-add using the Source-Extractor background method available in the {\sc photutils} python package \citep{larry_bradley_2025_14889440} by setting an unfiltered box size of 32 pixels and a 3$\sigma$ clipping algorithm.
Aperture photometry has been performed on each background subtracted co-added field at the position of M101-OT, and at multiple field stars, using an aperture radius of 2\,pixels ($\sim$5.5{\arcsec}). The photometry was calibrated by matching the field stars to the \texttt{AllWise} catalogue.
To obtain the final flux of M101-OT, we removed the host-galaxy contamination by subtracting the flux at the position of M101-OT of the earliest co-added images (non-detection; taking as baseline the median of the flux measured at 55356 and 55545 MJD epochs) from the flux at the same position obtained at later epochs.
The calculated magnitudes are listed in Table\,\ref{tab:photometry_table} of Appendix\,\ref{ap:photometry_table} and are shown in Fig\,\ref{fig:figure1}, which also includes the optical light curve evolution, extracted from \citet{2017ApJ...834..107B}, as reference, and taking $t=0$ at the first $r$-band peak (MJD 56970).

\subsection{Spitzer data}

The Spitzer Space telescope observed M101-OT from March 2004 to August 2019  with the NIR Array Camera \citep[IRAC;][]{2004ApJS..154...10F} in the \textit{[3.6]} and\textit{ [4.5]}\,{\micron} bands (\textit{I1} and \textit{I2}, respectively) as part of the SPitzer InfraRed Intensive Transients Survey (SPIRITS) programme \citep{2017ApJ...839...88K}.
To extract the photometry of M101-OT, we downloaded all images from the NASA/IPAC IRSA Spitzer Heritage Archive\footnote{\url{https://irsa.ipac.caltech.edu/applications/Spitzer/SHA}} (SHA).
Aperture photometry was performed on each Spitzer image at the position of M101-OT.
Due to the high resolution of Spitzer images ($\sim$0.6 {\arcsec}\,pix$^{-1}$), we calculated the background in an annulus (using a 3$\sigma$ clipping) and subtracted it from the aperture flux. 
Finally, we applied aperture correction following the Spitzer IRAC Instrument Handbook recipes\footnote{\url{https://irsa.ipac.caltech.edu/data/SPITZER/docs/irac/iracinstrumenthandbook/1/}} (Table 4.7), setting the aperture radius to 3\,pix and an annulus of 4$-$7\,pix.
Following the same methodology as in NEOWISE, to remove any residual flux from the host-galaxy, we also subtract the median baseline flux from epochs taken at MJD 52072 (March 2004).
The calculated magnitudes are shown in Fig.\,\ref{fig:figure1} (star and cross marks for I1 and I2 bands, respectively) and are listed in Table\,\ref{tab:photometry_table} in the Appendix\,\ref{ap:photometry_table}.

\subsection{NIRC2 data}

M101-OT was observed with the Laser Guide Star Adaptive Optics (LGSAO) available for the NIR camera \citep[NIRC2;][]{2000PASP..112..315W} mounted on the Keck II telescope located at Maunakea (Hawaii).
The observations took place on 6 August 2016, 4 April 2017, 4 July 2018, and 21 January 2019 in the \textit{J}, \textit{H} and \textit{Ks} bands as part of a LRN remnant monitoring programme (PI: Blagorodnova).
All observations were obtained in the wide camera mode, covering a field of view of 40{\arcsec}$\times$40{\arcsec}, with a pixel size of 0.04{\arcsec}\,pix$^{-1}$.

\begin{figure}
    \centering
    \includegraphics[width=1.0\linewidth]{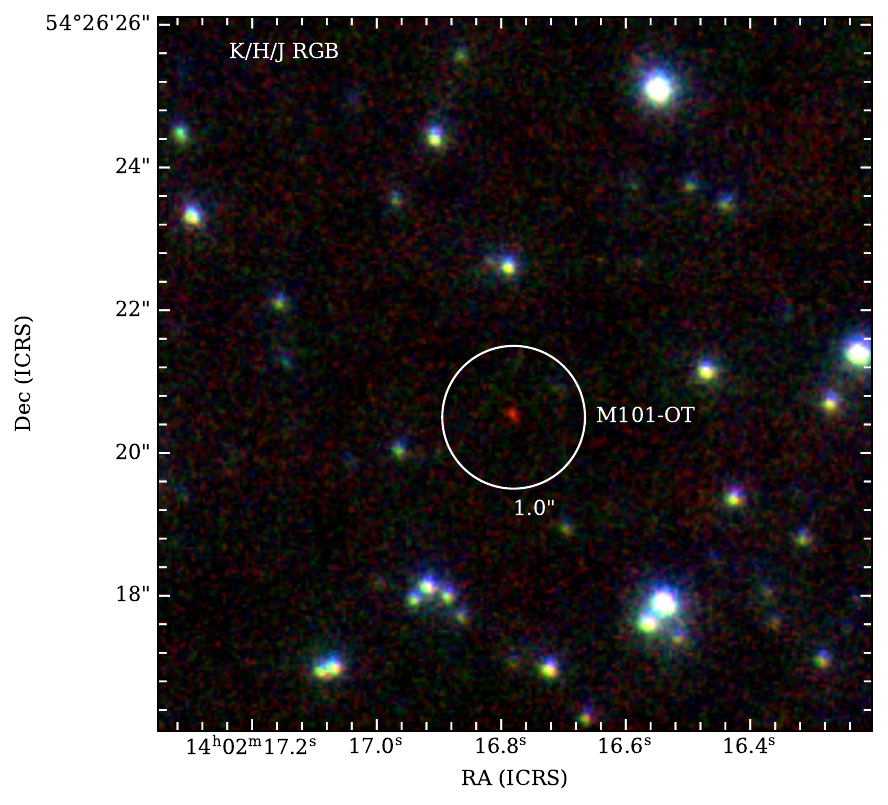}
    \caption{Colour composite of J (blue), H (green), and K (red) bands of the LRN M101-OT for the latest epoch
    obtained with the Adaptive Optics NIRC2 instrument (+1746\,d). The white circle indicates the position of M101-OT. \label{fig:figure2}}
\end{figure}

Images were reduced using custom-developed \texttt{python} routines that include dark current and flat calibrations.
We also removed the sky contribution in each band by subtracting a median-combined image taken at empty sky positions close to M101-OT before and after science observations.
A colour composite of the NIRC2 AO data of M101-OT is shown in Fig.~\ref{fig:figure2}. Provided its isolated location, the fluxes for M101-OT were measured using aperture photometry in the NIR.
The fluxes were calibrated using dedicated HST IR standard stars taken during the observing night for the dates of 6 August 2016 and 4 April 2017, while for the 4 July 2018 and 21 January 2019, we used the measured magnitudes of field stars calibrated from previous epochs. The observation log is listed in Table~\ref{tab:log_obs}.

\subsection{MOSFIRE data}

Follow-up imaging and spectra for M101-OT were taken with the MOSFIRE \citep{2012SPIE.8446E..0JM} photometric camera and spectrograph on Keck II as part of SPIRITS programme.
The imaging observations took place on 30 May 2016 and 8 October 2018 (PI Kasliwal) in the \textit{K} band (covering a field of view of 6.1{\arcmin}$\times$6.1{\arcmin} with a pixel size of 0.17{\arcsec}\,pix$^{-1}$), whereas the
spectra were taken on 31 March and 25 May 2015 (PI Prince), and 16 April 2016 (PI Kasliwal) using a slit-width of 0.7{\arcsec}.
We extracted the $K$ band fluxes using aperture photometry on the position of M101-OT and calibrated against the 2MASS catalogue using field stars.
The spectroscopic observations were taken in the NIR \textit{YJ}, \textit{H} and \textit{K} bands, and were reduced
using the \texttt{MosfireDRP} pipeline\footnote{Originally developed by N. Konidaris and C. Steidel at Caltech.
See \url{https://keck-datareductionpipelines.github.io/MosfireDRP/} .}. All the spectra were corrected by redshift (see Table~\ref{tab:m101_properties}). The log of the observations is listed in Table~\ref{tab:log_obs}.

\subsection{Gaia DR3 data}
The Gaia satellite \citep{2016A&A...595A...1G} serendipitously observed the region containing the  LRN M101-OT from 31 July 2014 to 10 May 2015, covering just before and after of the $r$-band first peak.
We retrieved the \textit{G}, \textit{G$_{RP}$} and \textit{G$_{BP}$} multi-epoch photometry
by searching for M101-OT in the Gaia DR3 epoch photometry table \citep{2023A&A...674A...1G};
we found that the Gaia DR3 1609279651165887104 correspond to M101-OT ($\sim$0.2{\arcsec}
within the coordinates listed in Table~\ref{tab:m101_properties}).
The Gaia multi-epoch photometry is shown in Fig.~\ref{fig:figure1}
and is also listed in Table~\ref{tab:photometry_table}.

\section{Photometric analysis}

In this section, we analyse the late-time light curve evolution of the LRN M101-OT by combining photometric measurements from the literature \citep{2017ApJ...834..107B}, archival photometry from Gaia DR3,  and new photometric follow-up in the NIR (see Sect.~\ref{sec:obs_data}). Although some of the NEOWISE and Spitzer photometry were already published in \cite{Reguitti2026}, we independently reduced and analysed the data, providing a point of comparison to their study (see Appendix~\ref{ap:mir_photometry_comparison}).

\begin{figure*}
    \centering
    \includegraphics[width=0.85\linewidth]{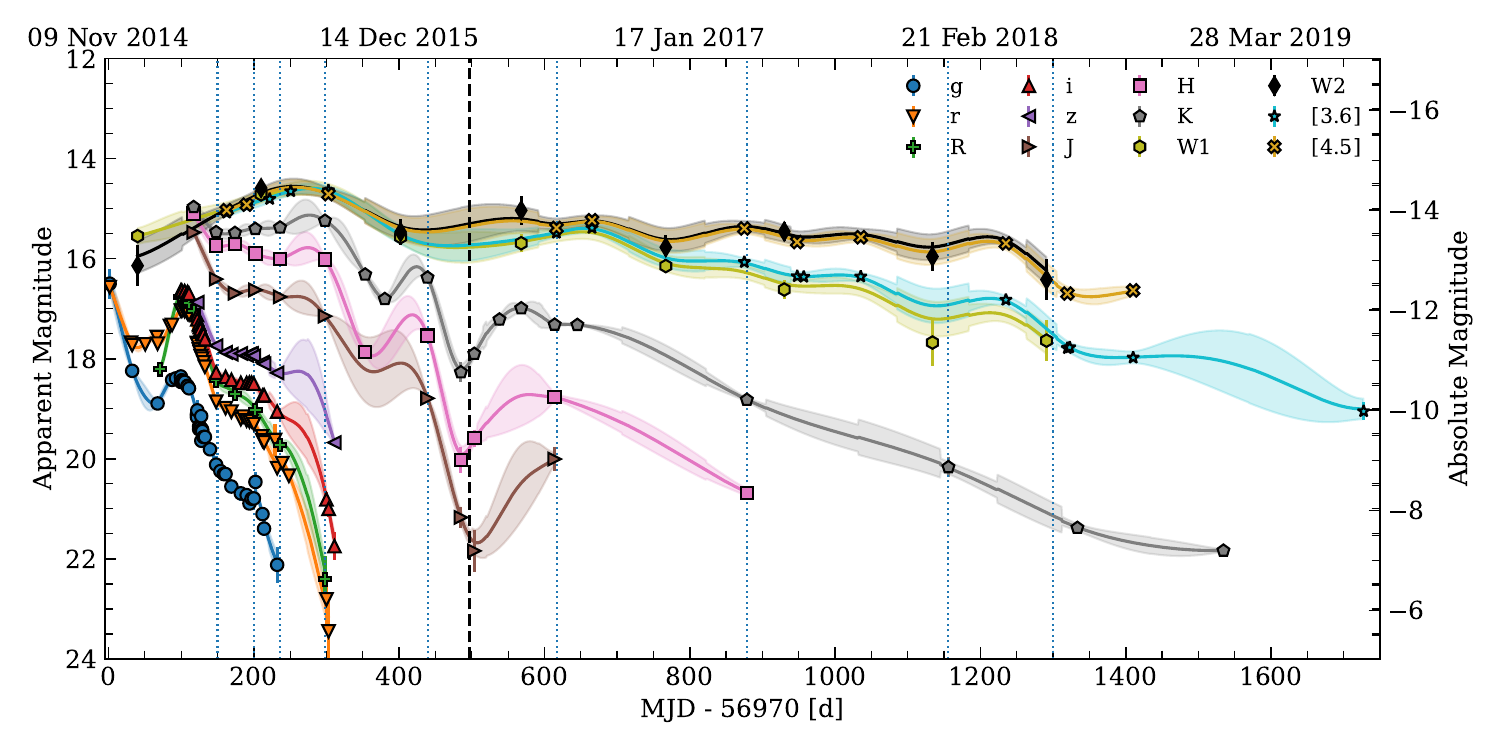}
    \caption{Interpolated multi-band light curve of the LRN M101-OT resulting from the multi-dimensional Gaussian process fitting (solid lines).  Shaded regions indicate the associated 1$\sigma$ errors. The vertical dashed line at $\sim$500\,d marks the transition to the late-time NIR re-brightening phase, while the dotted vertical lines represent the extracted SEDs. \label{fig:figure3} 
    }
\end{figure*}

\subsection{Colour evolution}
\label{subsec:color_evolution}

The multi-band light curve of M101-OT is shown in Fig.\,\ref{fig:figure1}.
The early-time optical evolution exhibits the characteristic double-peaked profile of a LRN:
an initial peak at t=0\,d (MJD 56970; taking the $r$-band as reference),
followed by a second maximum at $\sim$100\,d.
The second peak is followed by a brief decline leading into a short plateau in the redder bands between $\sim$140\,d and 160\,d,
before a final optical decline sets in \citep[at $\geq$160\,d; see][]{2017ApJ...834..107B}.

The late-time NIR evolution is marked by a re-brightening peaking at $\sim$300\,d, ($M_{K}=-13.68$)
followed by a decline to a local minimum at $\sim$500\,d ($M_{K}=-10.76$).
Subsequently, a second, lower-amplitude re-brightening
appears at 600\,d ($M_{K}=-12.04$) before fading up to $M_{K}=-7.20$ at $\sim$1550\,d.
In the MIR, the emission peaks at $\sim$300\,d and then
declines at a wavelength-dependent rate; the $W1$ (\textit{[3.6]}) band fades more rapidly ($2.3\times10^{-3}$\,mag\,day$^{-1}$) than the \textit{W2} (\textit{[4.5]}) band ($1.1\times10^{-3}$\,mag\,day$^{-1}$),
with the latter exhibiting near-plateau behaviour during this epoch.

To analyse the colour evolution of M101-OT, we interpolated the multi-band photometry using a
two-dimensional Gaussian Process (GP) regression implemented in our custom-developed \texttt{Python} package \texttt{MergerCurve} \citep[][DOI: \href{https://doi.org/10.5281/zenodo.20795378}{10.5281/zenodo.20795378}]{gomez_munoz_2026_20795378}. Our approach expands on the methodology used in the transient light curve classification software
{\sc Avocado}\footnote{\url{https://github.com/kboone/avocado}}\citep{2019AJ....158..257B}.
Similarly, our package also uses a 2-dimensional Matern\,3/2 kernel for the time of observation and wavelength of each data point.
In addition, our implementation anchors the central wavelengths separately across the optical, NIR, and MIR regimes
to better capture the distinct evolution in each range. For example, {\sc{Avocado}} was specifically designed for the classification of optical transient light curves from the Vera Rubin's Legacy Survey of Space and Time \citep[LSST;][]{2019ApJ...873..111I}, so the evolution across wavelengths was quite uniform. However, the extreme colour evolution observed in LRNe, requires a split into these different wavelength ranges.
Furthermore, to enforce the physical constraint of positive semi-definite flux and handle the high
dynamic range of the outburst, we performed the GP regression in log-flux space.
To better sample the GP regression across the first peak, we converted the Gaia DR3 photometry to $r$ and $g$  bands following the Gaia DR3 documentation\footnote{\url{https://gea.esac.esa.int/archive/documentation/GDR3/Data_processing/chap_cu5pho/cu5pho_sec_photSystem/cu5pho_ssec_photRelations.html}}. We also retrieve the $VR$ photometry from \citet{Goranskij2016AstBu} and applied the \citet{2006A&A...460..339J} transformation to derive the
corresponding $g$-band value at the first peak.
The resulting interpolated light curves are shown in Fig.\,\ref{fig:figure3}.

\begin{figure}
    \centering
     \includegraphics[width=0.98\linewidth]{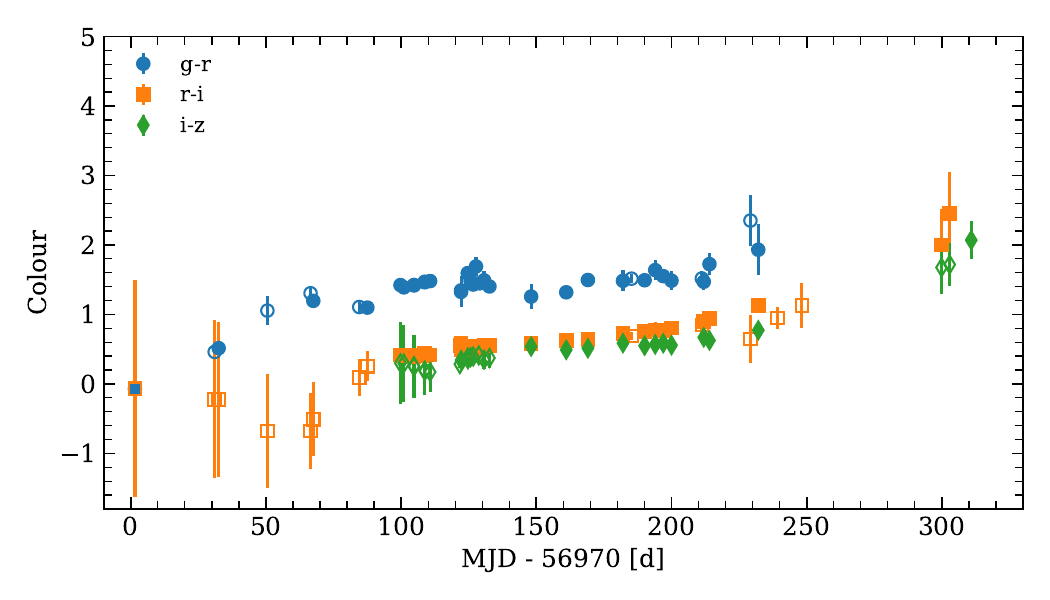}
    \includegraphics[width=0.98\linewidth]{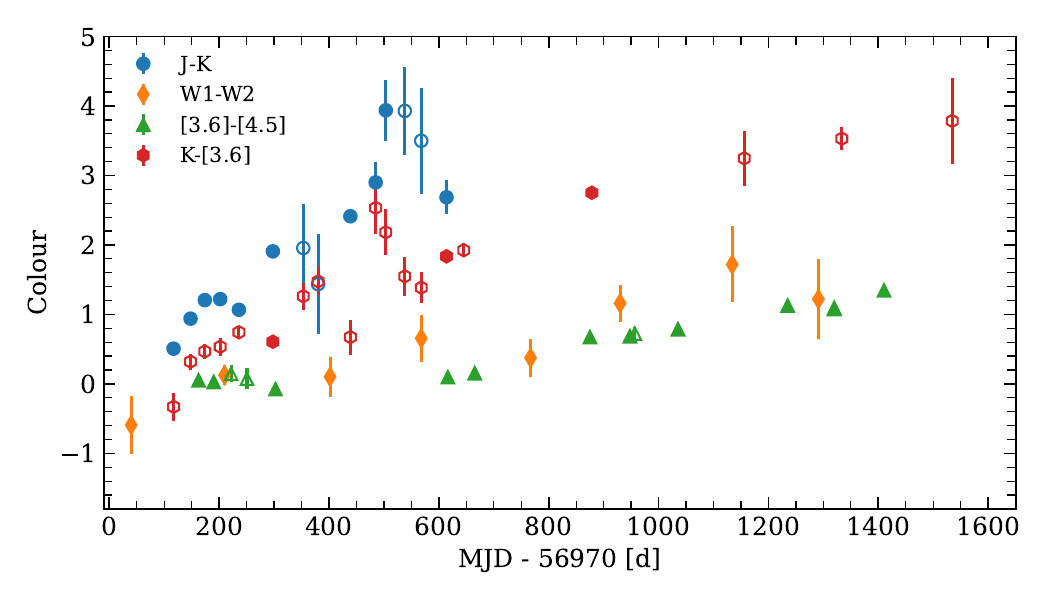}
    \caption{Post-outburst colour evolution of M101-OT in the optical (top panel) and IR (bottom panel) bands. Open markers denote colours for which one of the constituent bands was derived from the GP interpolation. \label{fig:figure4}}
\end{figure}

The optical colour evolution of the transient is shown in top panel of Fig.\,\ref{fig:figure4}. In comparison with the colour during the first peak ($g-r\simeq-0.1$), for which photometry is limited, the colours during the second peak appear much redder, with $g - r \simeq 1.4$\,mag, after that show a general redward trend over time, consistent with the cooling of the photosphere, which is evolving toward an M-type like star.

The IR colour evolution (bottom panel of Fig.\,\ref{fig:figure4}) shows three distinct features.
First, the NIR and NIR-MIR, in the $J-K$ and the $K-$\textit{[3.6]} colours, show a steady increase from $\sim$0 to $\sim$500\,d, similar to the early-time evolution seen in the $J-K$ colour in \citet{2017ApJ...834..107B}. Then, the colours reach a prominent red peak at $\sim$500\,d. This peak
occurs when the transient reaches its deepest photometric minimum, just prior to the late-time
NIR re-brightening observed at $\geq$500\,d. After the re-brightening, the colour evolution shows a redward trend over time. Simultaneously, the MIR colours ($W1-W2$ and \textit{[3.6]$-$[4.5]})
show a steady increase from $\sim$0 to $\sim$2\,mag over the light curve late-time evolution.
This persistent MIR reddening, combined with the slow luminosity decline, provides strong
evidence of the presence of cooling dust within the expanding material (see Sect.~\ref{subsec:sed_evolution}).
We note that there is a difference between our MIR colours and those presented in \citet{Reguitti2026} due to systematic shift in the $W1$ and $[3.6]$ (see Appendix~\ref{ap:mir_photometry_comparison}). However, in both datasets, the physical trend of gradual late-time reddening is consistently observed across both studies.

\subsection{SED evolution}
\label{subsec:sed_evolution}

To estimate the physical properties of the dust, we analysed the post-outbust SEDs of M101-OT at various epochs using
the interpolated multi-band light curve (see vertical dotted lines in Fig.\,\ref{fig:figure3}).
Specifically, we focused on the epochs that provided the best constraints,
selecting only those SEDs where observations covered a broad range of wavelengths
(see Table\,\ref{tab:dusty_models}).

The SED fitting was carried out using the custom-developed open source {\sc pyDusty} python
package\footnote{\url{https://github.com/mgomezAstro/pyDusty}}
\citep[a wrapper to handle the radiative transfer code {\sc DUSTY}\,v4;][]{1997MNRAS.287..799I,1999ascl.soft11001I,2000ASPC..196...77N}.
To model the observed SED, we used both a blackbody component only and a dust component, using the
\texttt{BBModel} and the \texttt{DustyModel}, respectively, available in {\sc pyDusty}.
The fitting process employs the method \texttt{EmceeRunner} that uses the {\sc emcee}
package for Markov-Chain Monte-Carlo \citep[MCMC;][]{2013PASP..125..306F}.
Both models assumed a uniform prior on the effective temperature, $T_\mathrm{eff}$, 
between 2000\,K and 15000\,K (extended down to 500\,K for the \texttt{BBModel}).
For the \texttt{BBModel}, the bolometric luminosity, $\log L/L_{\sun}$, is included as an explicit free parameter.
For the \texttt{DustyModel}, we assume a silicate dust composition \citep{1984ApJ...285...89D} with a standard ISM grain size distribution \citep{1977ApJ...217..425M}, a relative shell thickness $Y \equiv R_\mathrm{out}/R_\mathrm{in} = 2$, where $R_\mathrm{in}$ and $R_\mathrm{out}$ are the inner and outer shell radii, respectively, and a density profile $\rho \propto r^{-p}$ with $p=1$, representing the density profile generally adopted for supernova ejecta \citep[e.g.][]{1989ApJ...341..867C,2025ApJ...994L..41K}.

\begin{figure*}
    \centering
    \includegraphics[width=1.0\linewidth]{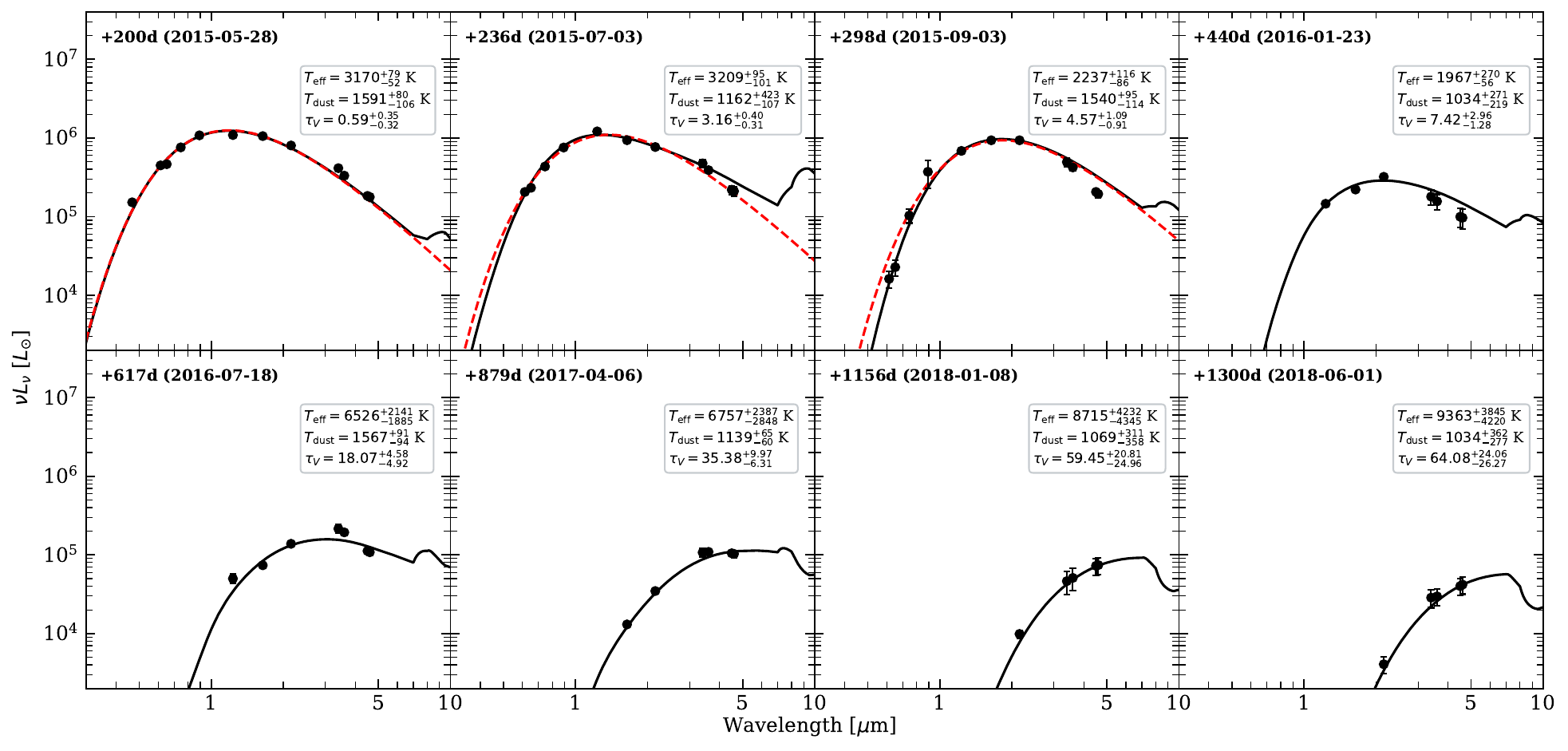}
    \caption{Evolution of the SED of M101-OT at different post-outburst peak epochs (relative to the MJD 56970). Fluxes were extracted from the GP processes described in Sect.~\ref{subsec:color_evolution}, already corrected for foreground extinction $E(B-V)=0.008$. The black solid line shows the best-fitting {\sc pyDusty} model while the
    red dashed line indicates a blackbody fit.   \label{fig:figure5}}
\end{figure*}

The MCMC sampler for \texttt{DustyModel} explicitly optimises three free parameters with uniform priors, including $T_\mathrm{eff}$, the visual optical depth, $\tau_V \in [0, 100]$, and the dust temperature at the inner boundary, $T_\mathrm{dust} \in [150\,\mathrm{K}, 1700\,\mathrm{K}]$.
In this case, the $\log L/L_{\sun}$ is not treated as an independent MCMC parameter, instead, it is determined analytically at each step by solving for the overall flux scale via weighted linear least squares. The inner dust shell radius, $R_\mathrm{in}$, is then derived post hoc by rescaling the default normalised DUSTY output ($L=10^{4}L_{\sun}$) to the fitted bolometric luminosity.
Then, to calculate the dust mass, $M_\mathrm{dust}$, at each epoch, we follow the formulation described in \citet{2025ApJ...983...87L} but for $p=1$, obtaining,
\begin{equation}
	\label{eq:dust_mass}
	\displaystyle \frac{M_\mathrm{dust}}{M_\sun} \approx 1.98841\times10^{-33} \left( \frac{2 \pi \tau_V R^{2}_\mathrm{in}}{\kappa_V} \right) \left( \frac{Y^{2} - 1}{\ln Y} \right)
\end{equation}
where $\kappa_{V}$ is the optical dust opacity per unit mass. Using the Mie theory \citep{1998asls.book.....B}, we
calculated a value of $\sim1.1\times10^{4}$\,cm$^{2}$\,g$^{-1}$ for the silicate grains. 
The best-fit models of the different epochs are shown in Fig\,\ref{fig:figure5}, with
parameters and uncertainties listed in Table\,\ref{tab:dusty_models}.

At the earliest modelled epoch (+150\,d), during the decay from the second peak, the SED is successfully reproduced by a purely
stellar blackbody with no evidence of dust obscuration.
At this stage, the central source resembles a luminous supergiant-like object with an
$T_\mathrm{eff}=3300$\,K, a $\log L/L_{\sun}=6.28$, and an extended stellar photosphere
of $\log R_{*}/\textrm{cm}=14.46$ ($\simeq$ 4150\,\Rsun).

Circumstellar dust formation begins rapidly at day +200, during the plateau, initially at a low optical depth shell
($\tau_V = 0.6$) with a high initial dust temperature ($T_\mathrm{dust} = 1591\,\text{K}$) and $\log R_\mathrm{in}/\text{cm} = 14.96$.
Over the subsequent 240 days, after the end of the plateau, the shell enters a strong dust condensation phase.
The $\tau_V$ increases by a factor of 20, peaking at
$\tau_V = 7.4$ at +440\,d, in which a total
dust mass of $\log M_\mathrm{dust}/M_\odot \sim -5.01$ is derived. During this period, as the dust shell condenses around the system,
the underlying stellar remnant remains cool ($T_\mathrm{eff} \sim 2000-3100$\,K).

We note that at early post-outburst epochs ($t \lesssim 236~\mathrm{d}$), the optical and NIR SED
can be reproduced by simple blackbody fits, reflecting a regime of low circumstellar extinction 
($\tau_V \approx 0.6$). However, to maintain a consistent physical framework across the entire
observational baseline, we modelled all epochs uniformly using the radiative transfer code
\textsc{DUSTY}. This enables us to self-consistently trace the continuous temporal evolution
of the circumstellar envelope, specifically the $\tau_V$, the 
$R_{\mathrm{in}}$, and the cumulative $M_{\mathrm{dust}}$, as the system transitions from an 
optically thin state to optically thick, heavily dust-obscured phase.

At epoch +500\, days, there is a minimum deep seen in the
\textit{JHKs} bands light curve (see Fig.~\ref{fig:figure3}). After this point, by +617\,d, the $T_\mathrm{eff}$ of the central source increases up sharply to 
$6526$\,K, eventually reaching $9363$\,K at +1300\,d.
At the same time, the stellar photosphere contracts monotonically by nearly two orders of magnitude,
dropping from a maximum of $\log R_*/\text{cm} = 14.73$ at +298\,d down to
$\log R_*/\text{cm} = 12.89$ at +1300\,d (at a contraction velocity of $\sim 61$\,{\kms}).

Parallel to the heating event of the central remnant, the $T_\mathrm{dust}$ exhibits an initial
decline from 1591\,K at +200\,d down to 1034\,K at +440\,d, followed by a re-heating phase reaching 1567\,K at 617\,d, before undergoing a second steady decline to 1034\,K at 1300\,d.
This $T_\mathrm{dust}$ bump at +617\,d is 
also evidenced by the NIR re-brightening phase observed at the same epochs (see Sect.~\ref{subsec:color_evolution}).
After the re-brightening phase, the optical depth increases substantially from
$\tau_{V} = 18.1$ to $64.1$, which is consistent with the increase in the dust temperature and
triggering dust formation.
Following this, the system transition into a period of steady dust condensation that sustains the MIR emission up to +3.55 years (reaching $\log M_\mathrm{dust}/M_{\sun} = -3.65$, $T_\mathrm{dust}=1034$\,K, and $\tau_{V} = 64.1$).

\begin{table*}
    \renewcommand{\arraystretch}{1.2} 
    \centering
    \caption{Posterior parameters from {\sc pyDusty} MCMC models of the LRN M101-OT remnant SEDs. The reference epoch for computing the phases are the first-peak outburst in the $r$-band at MJD 56970. \label{tab:dusty_models}}
    \begin{tabular}{cccccccccc}
    \hline
    \hline
        Date & Epoch & $\log L/L_{\sun}$ & $T_\mathrm{eff}$ & $\log R_\mathrm{*}$\tablefootmark{a} & $\tau_{V}$ & $T_\mathrm{dust}$ & $\log R_\mathrm{in}$ & $\log M_\mathrm{dust}/M_{\sun}$ & $\chi^{2}_\mathrm{min}$ \\
         & (d) &  & (K) & (cm) &  & (K) & (cm) &  & \\
    \hline
        2015-04-08 & 150 & 6.28$^{+0.01}_{-0.01}$ & 3330$^{+18}_{-17}$ & 14.46$^{+0.01}_{-0.01}$ & .\,.\,. & .\,.\,. & .\,.\,. & .\,.\,. & 50$^{+3}_{-1}$ \\
        2015-05-28 & $200$ & $6.23\pm^{0.00}_{0.00}$ & $3170\pm^{79}_{52}$ & $14.48\pm^{0.01}_{0.02}$ & $0.6\pm^{0.3}_{0.3}$ & $1591\pm^{80}_{106}$ & $14.96\pm^{0.08}_{0.05}$ & $-6.23\pm^{0.30}_{0.41}$ & $55\pm^{3}_{2}$ \\
        2015-07-03 & 236 & $6.24\pm^{0.02}_{0.03}$ & $3209\pm^{95}_{101}$ & $14.47\pm^{0.02}_{0.02}$ & $3.2\pm^{0.4}_{0.3}$ & $1162\pm^{423}_{107}$ & $15.35\pm^{0.11}_{0.30}$ & $-4.73\pm^{0.27}_{0.60}$ & $89\pm^{3}_{2}$ \\
        2015-09-03 & 298 & $6.12\pm^{0.01}_{0.01}$ & $2237\pm^{116}_{86}$ & $14.73\pm^{0.03}_{0.04}$ & $4.6\pm^{1.1}_{0.9}$ & $1540\pm^{95}_{114}$ & $14.92\pm^{0.07}_{0.07}$ & $-5.40\pm^{0.19}_{0.19}$ & $13\pm^{3}_{1}$ \\
        2016-01-23 & 440 & $5.65\pm^{0.02}_{0.02}$ & $1967\pm^{270}_{56}$ & $14.60\pm^{0.02}_{0.11}$ & $7.4\pm^{3.0}_{1.3}$ & $1034\pm^{271}_{219}$ & $15.01\pm^{0.19}_{0.14}$ & $-5.01\pm^{0.35}_{0.21}$ & $32\pm^{4}_{3}$ \\
        2016-07-18 & 617 & $5.44\pm^{0.04}_{0.03}$ & $6526\pm^{2141}_{1885}$ & $13.46\pm^{0.29}_{0.23}$ & $18.1\pm^{4.6}_{4.9}$ & $1567\pm^{91}_{94}$ & $15.03\pm^{0.12}_{0.10}$ & $-4.61\pm^{0.13}_{0.10}$ & $29\pm^{2}_{1}$ \\
        2017-04-06 & 879 & $5.28\pm^{0.07}_{0.05}$ & $6757\pm^{2387}_{2848}$ & $13.34\pm^{0.45}_{0.23}$ & $35.4\pm^{10.0}_{6.3}$ & $1139\pm^{65}_{60}$ & $15.26\pm^{0.13}_{0.12}$ & $-3.84\pm^{0.18}_{0.16}$ & $6\pm^{2}_{1}$ \\
        2018-01-08 & 1156 & $5.09\pm^{0.32}_{0.13}$ & $8715\pm^{4232}_{4345}$ & $13.08\pm^{0.56}_{0.34}$ & $59.5\pm^{20.8}_{25.0}$ & $1069\pm^{311}_{358}$ & $15.28\pm^{0.50}_{0.24}$ & $-3.57\pm^{0.74}_{0.40}$ & $43\pm^{3}_{1}$ \\
        2018-06-01 & 1300 & $4.87\pm^{0.26}_{0.15}$ & $9363\pm^{3845}_{4220}$ & $12.89\pm^{0.49}_{0.31}$ & $64.1\pm^{24.1}_{26.3}$ & $1034\pm^{362}_{277}$ & $15.23\pm^{0.40}_{0.28}$ & $-3.65\pm^{0.61}_{0.49}$ & $44\pm^{3}_{1}$ \\
    \hline
    \end{tabular}
    \tablefoot{
        \tablefoottext{a}{Calculated using $L=4\pi\sigma_\mathrm{SB}R_{*}^{2}T_\mathrm{eff}^{4}$, where $\sigma_\mathrm{SB}$ is the Stefan-Boltzmann constant.}
    }
\end{table*}

\section{Spectral analysis}

\subsection{NIR spectral evolution}

\begin{figure*}
    \centering
    \includegraphics[width=0.9\linewidth]{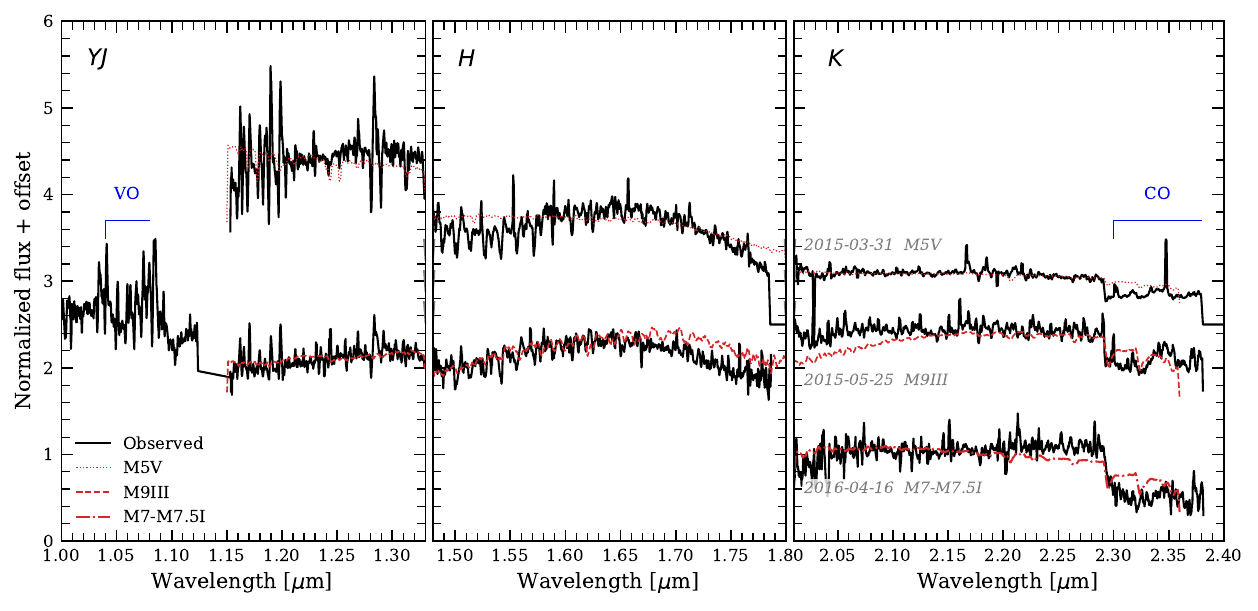}
    \caption{Comparison of M101-OT spectra with the best-fit spectral templates at +142, +197, and +524 (from top to bottom, respectively). The strongest molecular bands of VO and CO are marked in the figure, and are comparable with what it is observed for cool giant stars. Templates are for M5V, M9III, and M7$-$7.5I, as labelled in the figure, and are marked with dot, dashed, and dot-dashed lines, respectively. All the spectra were smoothed using a Gaussian kernel with FWHM=5\,{$\AA$} after $\sigma$-clipping to remove bad pixel and single-peak features for visualization purposes. \label{fig:figure6}}
\end{figure*}

Spectra taken at +142, +197, and +524\,d (after the second peak, during the plateau, and after the plateau, respectively), show the initial and progressive formation of
molecular bands, as in M-type stars.
Figure~\ref{fig:figure6} illustrates the spectral 
evolution of M101-OT at different NIR epochs, along with a comparison to best‑fit 
spectral types. We used the IR Telescope Facility (IRTF) spectral library
(SpeX; \citealt{2003PASP..115..362R}) and a $\chi^2$ minimization 
method to obtain the best‑matching templates; the resolution of the IRTF/SpeX templates was convolved to match the 
observed resolution of our data. The best matching spectral types correspond to M5V, M9III, and M7–M7.5I at +142, +197, and +524\,d, 
respectively, which agree with the cool central source temperatures derived from SED fitting  
($T_{\mathrm{eff}} = 3300$--$2050$\,K; see Sect.~\ref{subsec:sed_evolution}) and with the detection of TiO and VO in the optical spectra at similar epochs \citet{2017ApJ...834..107B}.
Note that these comparisons, on spectral types, serve primarily to constrain the temperature and continuum shape and they do not physically constrain the luminosity class of the progenitor system

\begin{figure*}
    \centering
    \includegraphics[width=0.8\linewidth]{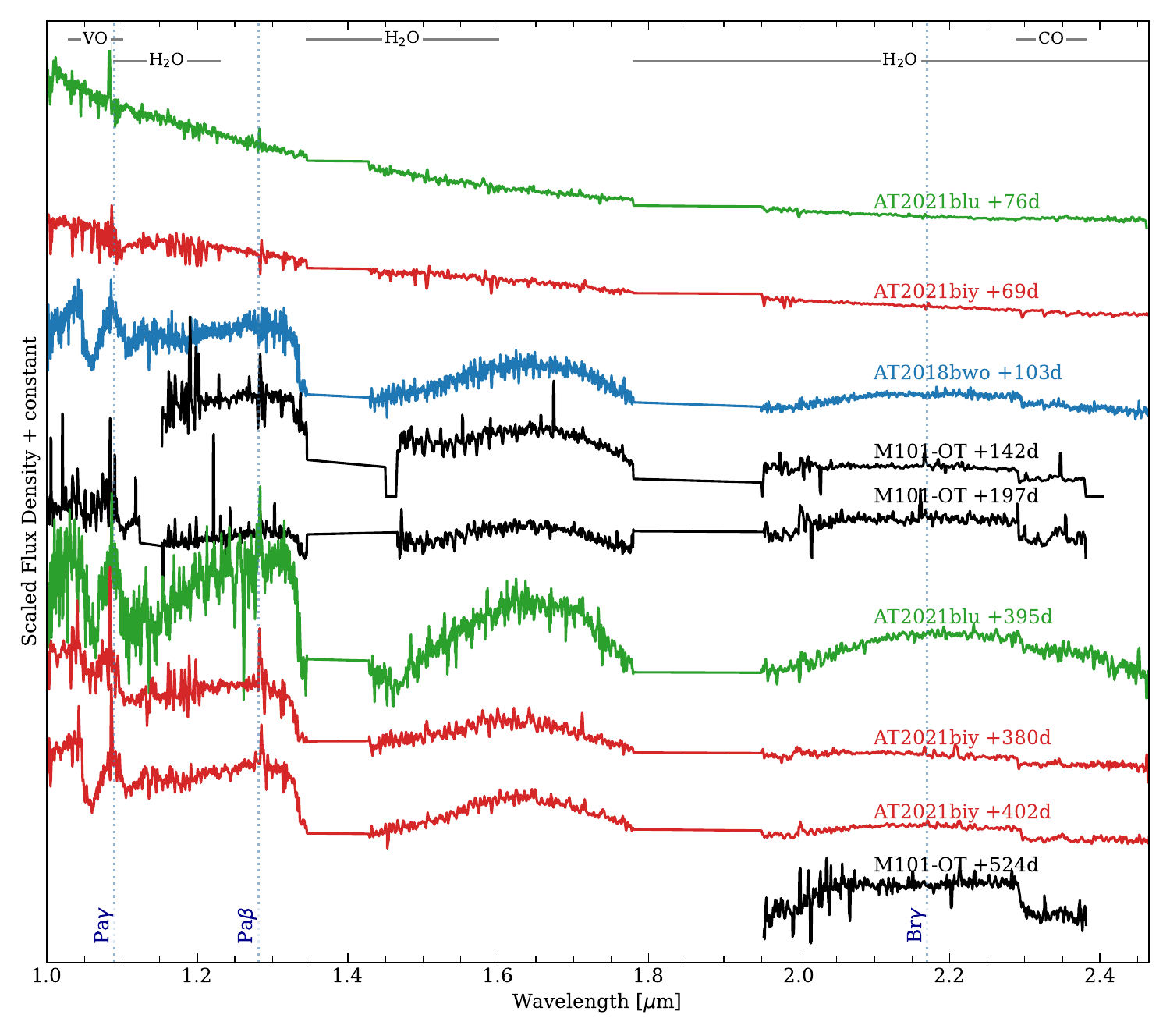}
    \caption{NIR evolution comparison between M101-OT (black solid lines) with respect to other LRNe (as labeled in the figure). The NIR spectra were extracted from \citet{2023ApJ...948..137K} and \citet{2021A&A...653A.134B}. The main molecular band regions are labelled in the figure. \label{fig:figure7}}
\end{figure*}

As a point of comparison, Fig.~\ref{fig:figure7} shows the NIR spectral evolution of the extragalactic LRNe AT\,2021blu, AT\,2021biy, and AT\,2018bwo, at various post-outburst epochs.
The spectra are compiled from \citet{2023ApJ...948..137K} and \citet{2021A&A...653A.134B}.
When compared to the M101-OT sequence, the different LRNe reveal a coherent evolutionary path
dominated by broad molecular absorption features.
These are primarily characterised by the electronic band $D--A$ (0,0) of VO (near $1.0\text{--}1.1\,\mu\text{m}$),
CO overtones, and broad water vapour bands
\citep[similar to those observed in V838~Mon; e.g.][]{2004ApJ...607..460L},
which are indicated at the top of the figure with horizontal lines.

The VO absorption feature seems to appear as early as $\sim$100\,d and remain persistent across different epochs. 
Similarly, the water absorption features emerge early in the evolution,
as they seem to appear by $\sim 70$\,d after the first post-outburst peak.
In contrast, the CO band head becomes prominent later in the sequence, appearing at $\sim 100$\,d and
strengthening significantly by $\sim 500$\,d (except for AT\,2021blu which signatures of CO absorption appears later in its evolution  at $\sim+395$\,d). While these molecules mirror the early-epoch NIR properties
of V838~Mon \citep{2004ApJ...607..460L}, they also align with the late-stage MIR molecular analysis recently
reported for AT~2021biy and AT~2021blu \citep{2026ApJ...999...16K}.

\subsection{Carbon-monoxide modelling}

We modelled the CO first overtone ($\Delta v = 2$) molecular absorption feature using an LTE isothermal slab model,
a standard approach for characterizing NIR molecular spectra.
The cross-sections were computed on a grid sampled every 10~K using line-by-line spectroscopic parameters
from the EXOMOL database \citep{2026MNRAS.tmp..926T}, including only the dominant $^{12}\text{C}^{16}\text{O}$
isotopologue \citep{2015ApJS..216...15L}. 
The model spectrum was then
convolved using a Gaussian kernel to match the observed spectra.
While MOSFIRE delivers a nominal instrumental resolution of $R \sim 3500$ in the $K$ band,
the resulting best-fit effective resolution of $R \sim 1000$ takes into account the combined effects
of the instrumental profile and the intrinsic macroscopic gas kinematics. This suggests intrinsic line widths of $\approx290$ km s$^{-1}$, but this value is highly uncertain due to poor instrumental resolution and the high-level of smearing in the band itself. The line width is mainly based on the shape of flux jump at the bandhead. 

\begin{figure}
    \centering
    \includegraphics[width=0.9\linewidth]{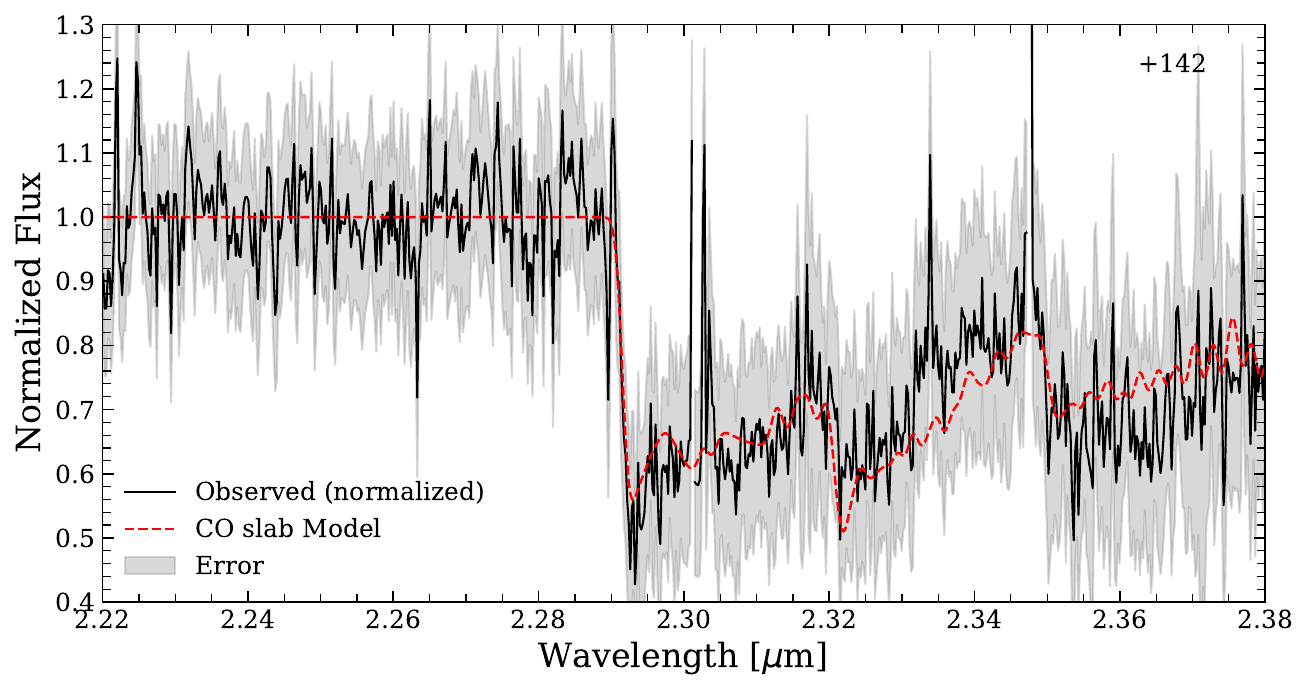}
    \includegraphics[width=0.9\linewidth]{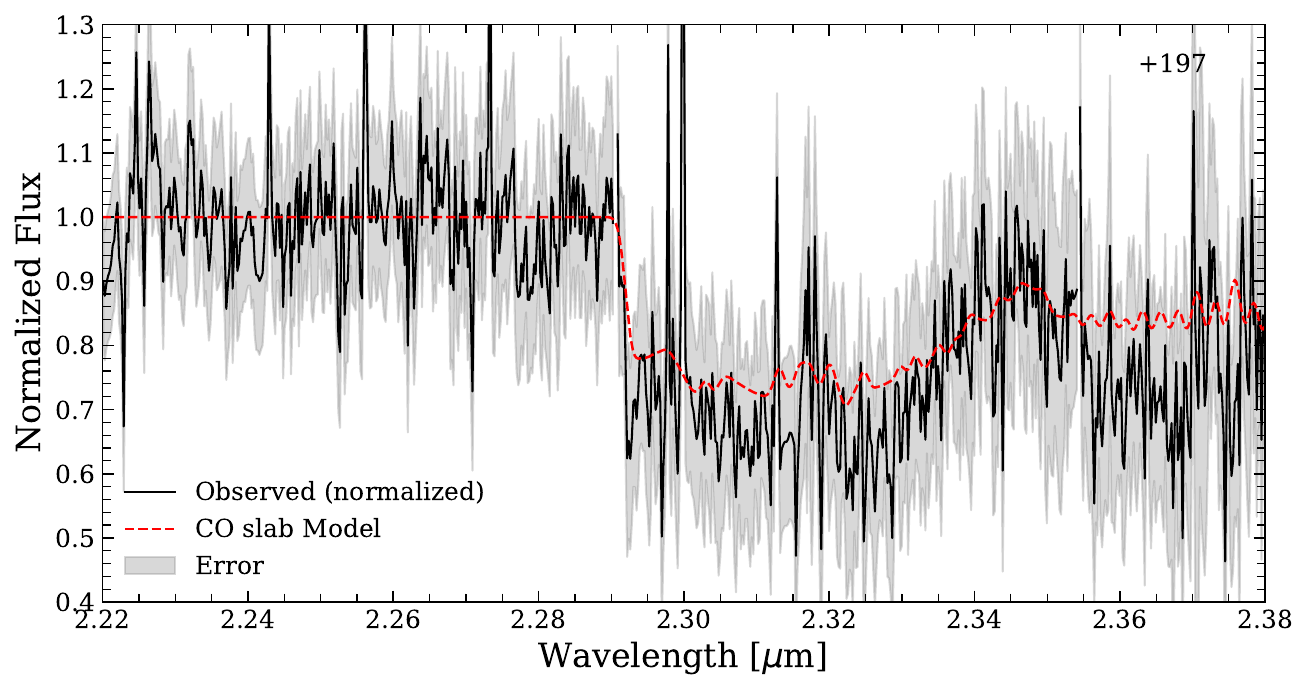}
    \caption{Comparison of M101-OT spectra (solid line) with the best-fit CO isothermal slab model (dashed line) for the epochs +142\,d (upper panel) and +197\,d (lower panel). The shaded region represent the observed spectral error. \label{fig:figure8}}
\end{figure}

Figure~\ref{fig:figure8} shows the best-fitting CO slab models overlaid on the observed MOSFIRE spectra.
At +142\,d, the CO first overtone absorption profile is best reproduced with a high CO column
density of $\log N_\mathrm{CO}/(\mathrm{cm}^{-2}) = 20.93$. Roughly 55 days later, at +197\,d, the column density
drops to $\log N_\mathrm{CO}/(\mathrm{cm}^{-2}) = 20.73$. 
Note that the CO features are located at the extreme edge of the spectral window introducing significant
uncertainty into the continuum normalization. 
Because the choice of the normalization polynomial is highly
degenerate with the derived gas temperature, the values of $T_\mathrm{gas}$ have large uncertainties,
with central values of 1900\,K and 1550\,K at +142 and +197\,d, respectively.
Conversely, the errors on $\log N_\mathrm{CO}$ are estimated to be within an order-of-magnitude. At our model parameters, the corresponding absorption in the second CO overtone ($\Delta v=3$, near 1.56 $\mu$m) is weaker than noise in the $H$-band spectrum.
Furthermore, the model resulted in a poor fit to the +524\,d epoch;
we suspect that water vapour may contaminate this spectral region,
although with the low signal-to-noise ratio (S/N) and resolution of the current data it could not be estimated reliably.

\subsection{NIR spectral line identification and line profiles}
\label{subsec:line_identifications}

Full NIR spectra of MOSFIRE observations over all three epochs are shown in Fig.~\ref{fig:C1}.
Across the $YJ$, $H$, and $K$ bands, the NIR spectra exhibit a rich mixture of atomic emission or absorption features alongside the broad molecular bandheads.
To identify candidate line transitions, we compared the observed NIR spectra against rest wavelengths retrieved from the NIST Atomic Spectra Database\footnote{\url{https://dx.doi.org/10.18434/T4W30F}}.
The identified species include \ion{Ti}{i}, \ion{Fe}{i}, \ion{He}{i}, and \ion{H}{i}.
In particular prominent \ion{Fe}{i} emission lines are detected in the $YJ$ band along with the strong emission of \ion{He}{i} at 1.0832\,{\micron}; the detection of \ion{Fe}{i} is consistent to what has been found in the optical spectra \citep[see e.g.][and spectra listed in WiseRep\footnote{\url{https://www.wiserep.org/object/1266}} for this object]{2017ApJ...834..107B}. 
The prominence of these cool, neutral metallic species is consistent with the late M-type spectral classifications and low effective temperatures ($T_{\mathrm{eff}} \lesssim 3300$\,K) resulted from the SED modelling in Sect.~\ref{subsec:sed_evolution}.

\begin{figure}
    \centering
    \includegraphics[width=0.9\linewidth]{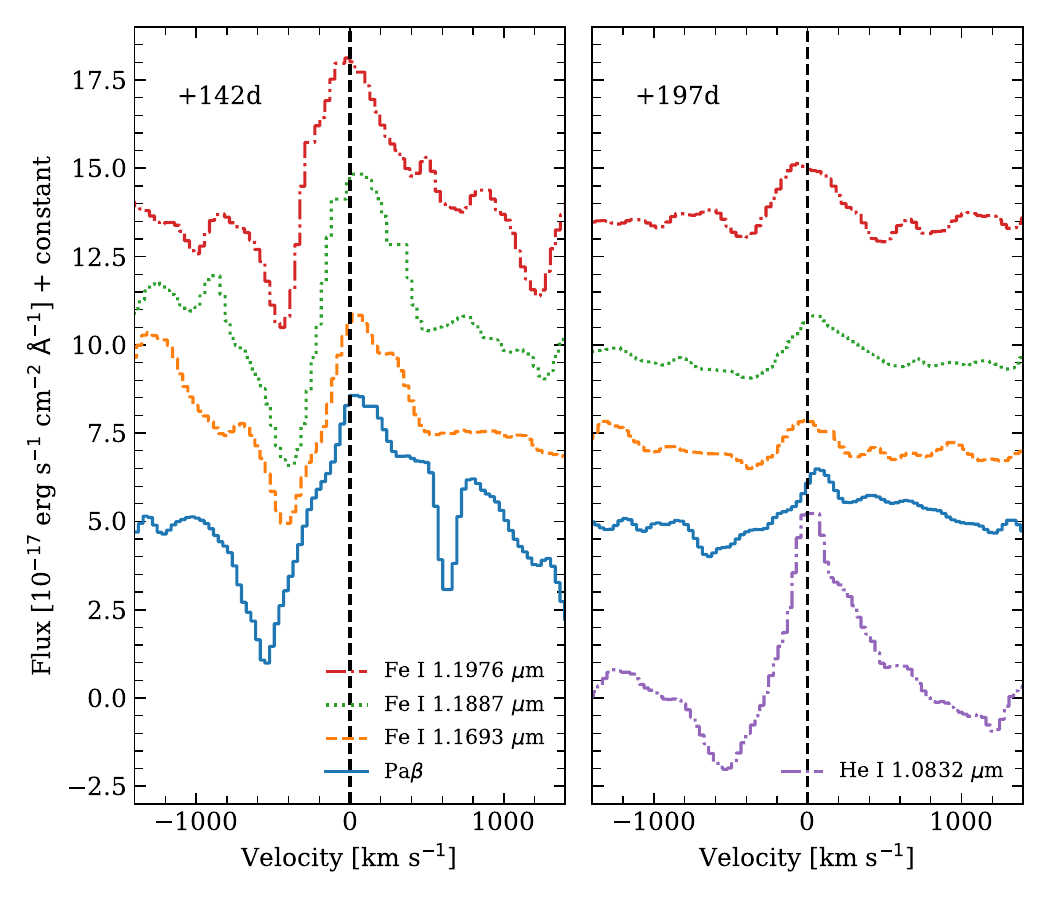}
    \caption{The profile of the Pa$\beta$ and \ion{Fe}{i} emission lines as seen in the NIR MOSFIRE spectra during the beginning (+142\,d; left panel) and during the plateau phase (+197\,d; right panel). The spectra were smoothed by a Gaussian kernel using a FWWHM=5\,{$\AA$} after applying a $\sigma$-clipping to remove bad pixels. For the second epoch, the \ion{He}{i}\,1.0832\,{\micron} is also shown. \label{fig:figure9}}
\end{figure}

In addition to these atomic emission features, Pa$\beta$ and the \ion{Fe}{i} multiplet exhibit distinct P-Cygni profiles, indicating the presence of expanding ejecta.
At the earliest observed epoch (+142\,d), which marks the onset of the plateau phase following the second peak, we measured expansion velocities, at the absorption minimum, of $\sim$400\,{\kms} for the \ion{Fe}{i} multiplet 60 lines (1.1690, 1.1884, and 1.1973\,{\micron}) and $\sim$520\,{\kms} for Pa$\beta$ (see Fig.~\ref{fig:figure9}).
By +197\,d, during the plateau phase, the \ion{Fe}{i} velocity had declined to $\sim$370\,{\kms}, whereas the Pa$\beta$ velocity increase to $\sim$620\,{\kms}; at this same epoch, the \ion{He}{i} emission line likely emerged, exhibiting an expansion velocity of $\sim$520\,{\kms}.
We note that the quoted \ion{Fe}{i} velocities represent averages calculated across the three individual lines of the multiplet.
The differences between the expansion velocity of the neutral species and high-excitation species trace the a radial velocity stratification, where metal lines trace the slower dense material while the H and He trace the faster and outer material.


\section{Discussion}

\subsection{Dust evolution in the LRN M101-OT}

The LRN M101-OT has shown an IR evolution characterised by the transition from a stable,
apparently dust-free progenitor to a heavily obscured, dusty remnant. In fact, archival data
from 15 to eight years prior to the outburst of M101-OT revealed an F-type yellow super-giant
progenitor with no evidence of a pre-existing warm dust component \citep{2017ApJ...834..107B}.
Following its double peak optical outburst, the SED of the transient shifted rapidly
toward longer wavelengths as the ejected material cooled. While the source faded below
detection limits in the optical bands during the first year (see Fig.~\ref{fig:figure1}),
it remained bright in the MIR for nearly five years
\citep[see][and our Fig.~\ref{fig:figure1}]{2026A&A...706A.154R}.
However, despite the detection of an IR excess indicative of dust condensation in
previous NIR \citep{2017ApJ...834..107B} and MIR \citep{2026A&A...706A.154R} observations,
 a detailed dust analysis had not been conducted.

\begin{figure}
    \centering
    \includegraphics[width=1.0\linewidth]{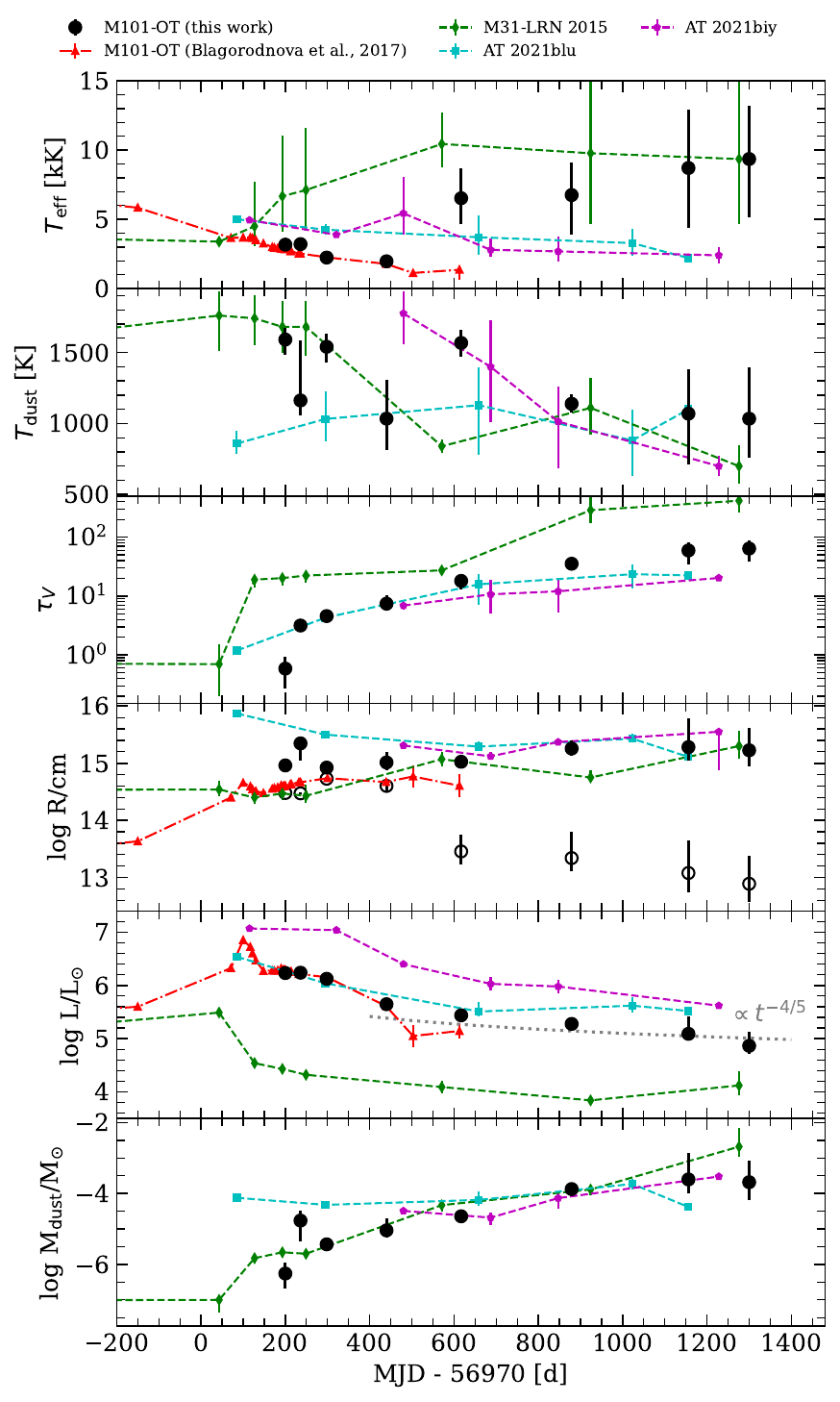}
    \caption{ Evolution of the dust properties of M101-OT as shown in Table~\ref{tab:dusty_models}. As reference, the evolution of the dust properties of M31-LRN-2015 \citep{2020MNRAS.496.5503B} and AT\,2021biy and AT\,2021blu \citep{2026arXiv260622619W} are also shown; the latest dust evolution properties for AT\,2021blu and AT\,2021biy were extracted from \citet{2026ApJ...999...16K}. Note that in the fourth panel (from top to bottom) the open symbols indicate the central remnant radius as described in Table~\ref{tab:dusty_models}. \label{fig:figure10}}
\end{figure}

The light curve evolution displayed in Fig.~\ref{fig:figure4} shows an IR excess
peaking at $\sim$500\,d after the outburst peak, followed by a decline
in the NIR colours, while remaining in a slow decline in the MIR colours (bottom panels).
By using GP interpolation of the optical and IR photometry, we were able to 
extract multiple SEDs at different epochs to study the dust evolution in M101-OT.
Figure~\ref{fig:figure10} shows the evolution of the stellar and dust shell
parameters of M101-OT, as obtained from the multi-epoch SED analysis presented in
Sect.~\ref{subsec:sed_evolution}.

At the earliest modelled epoch, at the beginning of the plateau after the second peak (at +150\,d), the SED was successfully reproduced by a simple
blackbody. However, optical spectra taken at similar epochs show signs of TiO and VO absorptions
\citep{2017ApJ...834..107B}.
This implies that if dust is present, it is optically thin, with a
$T_\mathrm{dust} \sim 1500$\,K and an upper limit of
$\log M_\mathrm{dust} \sim -6.36$\footnote{We use the \texttt{ThermalEmissionModel} included 
in {\sc pyDusty} to calculate the dust-mass upper limit. In this case, the free parameters are 
the $\log R/R_{\sun}$, $T_\mathrm{eff}$, and $\log M/M_{|\sun}$, for a fixed value of 
$T_\mathrm{dust}=1500$\,K, for silicate mineralogy, and a fixed grain size of 0.1\,{\micron}.}.
Such a component would not contribute significantly to the overall flux measured in the NIR,
but would provide the environment to allow for molecular formation.
This is supported by the evolution of the CO band. Between +142 and +197\,d, both the $\log N_\mathrm{CO}$ and $T_\mathrm{gas}$
decrease, with  $\log N_\mathrm{CO}$ declining from $-20.9$ to $-20.7$ and $T_\mathrm{gas}$ from 1900\,K to 1550\,K.
This decline in $N_\mathrm{CO}$ aligns with the expected geometric dilution and density drop of an expanding shell,
underlining the high density and survival of molecules during this phase of the LRN expansion \citep[see e.g.][]{2005ApJ...627L.141B,2018A&A...617A.129K}.
This material was likely ejected prior to the outburst peak and is possibly located far from the
stellar photosphere.
Signatures of pre-existing dust, during the progenitor phase and at the outburst peak, have been
found in other LRNe such as AT\,2021biy \citep{2026arXiv260622619W},
V1309\,Sco \citep{2016A&A...592A.134T}, and more recently in AT\,2025abao by
\citet{Reguitti2026} and Karambelkar et al. (in prep.; by early JWST observations).
Following this initial phase, our multi-epoch modelling reveals a transition into
a regime of continuous dust condensation sustaining the MIR emission for up to
five years. 

As shown in the second panel of Fig.~\ref{fig:figure10}, the dust temperature exhibits two distinct cooling declines separated by the NIR re-brightening observed in M101-OT (see Figs.~\ref{fig:figure1} and \ref{fig:figure4}). During the first decline, the dust temperature decreases from 1591\,K at 200\,d to 1034\,K at 440\,d, broadly resembling the early dust temperature evolution inferred for the low-mass progenitor LRN M31-LRN-2015.
In LRN M31-LRN-2015, the subsequent increase in the optical depth was interpreted as evidence that shocks enhanced dust formation \citep{2020MNRAS.496.5503B}.
Following the NIR re-brightening, the dust temperature increases again, reaching values of the condensation limits, resembling the evolution of AT\,2021biy afterwards. The DUSTY models also favour an increase in the temperature of the intrinsic illuminating source, reaching $T_\mathrm{*}\sim6500$\,K. 
However, as the dust optical depth increases, the central source becomes progressively obscured, making $T_\mathrm{*}$ degenerate with $\tau_V$ and consequently, this apparent increase should be treated with caution. 
The simultaneous re-brightening of the NIR emission (at $\sim$600\,d), re-heating of the dust, and rapid increase in the optical depth likely indicate a second phase of dust nucleation/reprocessing around the ejected material.
One possible explanation is shock interaction between faster ejecta and slower pre-existing circumstellar material \citep{2017MNRAS.471.3200M}, which can convert kinetic energy into radiation and heat the newly formed dust. The associated compression and rapid cooling of the post-shock gas may also enhance additional dust formation, providing a natural explanation for the subsequent increase in $\tau_{V}$ and dust mass after +617\,d (see the third and sixth panels of Fig.~\ref{fig:figure10}).
After the re-brightening, all LRNe shown in Fig.~\ref{fig:figure10} follow the same steady increase in dust mass and optical depth, with M101-OT reaching values of $\log M_\mathrm{dust}/M_{\sun}\sim-3.65$ and $\tau_{V}\sim64$.
These results suggest that late-time shock interaction and renewed dust formation may represent an important stage in the evolution of LRN remnants.

Regarding the bolometric luminosity (fifth panel of Fig.~\ref{fig:figure10}),
the late-time evolution follows a similar trend compared to other LRNe such M31-LRN-2015,
AT\,2021blu, and AT\,2021biy, namely, a peak and a steady decrease described by the gravitational contraction
\citep[$L \propto t^{-4/5}$; see][]{2005A&A...436.1009T,2020MNRAS.496.5503B}, with slight variations in luminosity as possibly signatures
of shock interactions such as the second bump present in AT\,2021blu and AT\,2021biy
(see Fig.~\ref{fig:figure10} between 600\,d and 1200\,d), as previously discussed for
a sample of LRNe in \citet{2026ApJ...999...16K}.
Additionally, all LRNe stabilize their luminosity
after $\sim400$\,d, except for M31-LRN-2015, that stabilizes at $\sim 200$\,d, marking a shift from early ejecta-dominated cooling to persistent heating driven by a possibly long-term shock interaction.

The fourth panel of Fig.~\ref{fig:figure10} shows the evolution of both the dust inner shell radius, $\log R_\mathrm{in}/\text{cm}$, and the evolution of the central remnant radius.
The expansion of the shell radius in M101-OT is consistent with the behaviour observed in similarly high-mass progenitor LRNe, such as AT\,2021blu and AT\,2021biy.
For these LRNe, \citet{2026ApJ...999...16K} inferred late-time expansion velocities of $\sim150$\,{\kms} and $\sim250$\,{\kms}, respectively, which closely aligns with the expansion velocity inferred for M101-OT by \citet{2017ApJ...834..107B} of $\sim550$\,{\kms} and our estimates for Pa$\beta$ emission line between $\sim520$\,{\kms} and $\sim620$\,{\kms}, at two different epochs. 
In contrast, the low-mass progenitor LRN M31-LRN\,2015 exhibits a different early-time evolution, showing a smaller shell radius.
As highlighted in \citet{KaminskiBlagorodnova2026}, lower-mass progenitors tend to have lower ejection velocities.

\subsection{Limitation of our models}

In Sect.~\ref{subsec:sed_evolution}, we modelled the multi-epoch SEDs using the radiative transfer code {\sc DUSTY}. For this modelling, we assumed a spherically symmetric shell containing pure silicate dust
with a standard MRN grain size distribution. However, various theoretical and observational studies
suggest the presence of significant non-spherical geometries, such as barrel-like, bipolar, or
circumstellar disk structures \citep[e.g.][]{2013A&A...558A..82K,2017MNRAS.471.3200M, 2018A&A...617A.129K, 2021A&A...655A..32K,2021AJ....162..183W}. 
Furthermore, dust grains in post-CE systems have been shown to deviate from standard distributions,
potentially following steeper power-law indices, such as $\alpha = -8.5$ \citep[e.g.][]{2024MNRAS.533..464B}, and hence slightly affecting the dust properties
derived in our models.

In addition, radiative-transfer models applied to LRNe typically assume a steady-state wind density distribution for the ejecta, $\rho \propto r^{-p}$, of $p=2$ \citep[e.g.][]{2020MNRAS.496.5503B,2026ApJ...999...16K,2026arXiv260622619W}. 
However, motivated by hydrodynamical and semi-analytic of stellar envelope ejection,
we assumed a shallower density distribution with $p=1$. 
Such profiles have been used to approximate dynamically ejected, shock-driven envelope material \citep{1989ApJ...341..867C,2025ApJ...994L..41K}, although the exact density structure remains uncertain.
Because a $p=1$ law concentrates a larger fraction of the total shell mass into the outer radii ($r > R_{\mathrm{in}}$), a higher total $M_{\mathrm{dust}}$ is required to achieve a given $\tau_V$ compared to a $p=2$ profile, which in our models translate into a higher $R_{\mathrm{in}}$ (see Eq.~\ref{eq:dust_mass}) for a fixed shell thickness.
In practice, for M101-OT, fitting the SEDs using $p=2$ resulted in higher values for $\tau_V$, lower values for of $T_\mathrm{dust}$, higher $R_{\mathrm{in}}$, and hence, higher values for the dust mass.
Note also that our models adopted the value for the thickness of the shell to be $Y=2$. However,
given the spectral coverage of our SEDs, our models are not sensitive to geometric variations, as
longer wavelengths would be needed (e.g. in the far-IR) in order to probe cooler dust at longer radii and
to better constrain the shell outer extent. 

Another point to consider is that molecular absorption may affect the reliability of the early-time SED fits.
Based on our MOSFIRE spectra, the SED at early epochs is influenced by molecular 
absorption, particularly from water and CO, implying the presence of the CO bandhead at 4.63\,{\micron},
and hence strongly deviated from a blackbody shape.
This spectral contamination may lead to a slight overestimation of the derived inner shell radius and,
consequently, the dust mass. This is particularly evident in the poor fit to the $W2$ band at +298\,d 
(see Fig.~\ref{fig:figure5}), where CO absorption features may be present \citep[see][]{2026ApJ...999...16K}. 
Finally, the assumption of a blackbody in our models as the central source may introduce additional uncertainties, particuarly at short wavelengths where cool stellar atmospheres can depart significantly from a Planck spectrum. Since the UV flux provide an important contribution to dust heating, deviation from a balckbody spectral shape could affect the inferred dust properties.
While these factors introduce systematic uncertainties, performing more 
sophisticated multi-dimensional radiative transfer modelling is beyond the scope of this work.

\section{Conclusions}

In this work, we presented a comprehensive late-time photometric and spectroscopic analysis of the M101-OT,
tracing its evolution up to $\sim$5 years after its initial 2015 outburst peak.

The early-time optical light curves display a LRN with a double-peaked profile followed by a brief
plateau phase between $\sim 140$ and $160$\,d, making it part of the `risers' light curve morphology. 
The redward trend in the optical colours ($g-r$, $r-i$, and $i-z$) tracks the
continuous cooling of the expanding photosphere toward an apparent M-type stellar 
classification, while  the NIR colours reveal a highly complex re-brightening profile,
represented by a peak at $ +617$\,d ($M_K = -12.04$) and preceded by a local minimum at +500\,d ($M_K \simeq -10.3$).
Finally, the steady, long-term reddening of the mid-infrared colours ($W1-W2$ and $[3.6]-[4.5]$)
from $\sim 0$ up to $\sim 2$\,mag, accompanied with slow decline in the MIR flux, 
may indicate a massive, cooling circumstellar dust surviving several years post-merger.

The progenitor of M101-OT is described in the literature as an F-type giant with an initial mass of $\sim$18\,M$_\sun$. Following the early
post-outburst evolution, between +150 to +524\,d, we corroborate that the central remnant is best described by a late M-type star according to our analysis
of the NIR spectra. This low-resolution MOSFIRE spectra confirmed an O-rich environment dominated by VO, water vapor, and CO overtones.
Modelling of the CO first-overtone bandhead ($\Delta \nu=2$) yielded high column densities
($\log N_{\rm CO}/\text{cm}^{-2} \approx 20.7$ and $20.9$ at +142\,d and +197\,d, respectively), similar to what was found for AT\,2021biy and 
AT\,2021blu at late epochs in their MIR evolution.

Circumstellar dust condensation initiated rapidly at +200\,d with an estimated $\log M_\mathrm{dust}/M_{\sun}=-6.23$. 
The MOSFIRE NIR spectra obtained at these epochs show prominent \ion{Fe}{i} and Pa$\beta$ P-Cygni profiles, with expansion velocities ranging between $\sim520$ and $\sim620$\,{\kms}.
The dust mass peaked at $\log M_{\rm dust}/M_\odot \approx -5.01$ around +440\,d, following a decline in the dust temperature
before the NIR re-brightening. After the NIR re-brightening, the dust continued a steady increase, up to $\log M_{\rm dust}/M_\odot \approx -3.65$ at the latest epoch (+1300\,d), following a second decline in the dust temperature.
The overall dust mass budget closely tracks that of high-mass progenitor LRNe such as AT\,2021blu and AT\,2021biy. However, the high columnar densities found in the CO bandhead at 2.3\,{\micron}
may indicate a saturation of the CO fundamental band at 4.67\,{\micron},
affecting the photometry measured in the $W2$ (and \textit{[4.5]}) band at early epochs and
hence, dust masses derived at those epochs are likely upper-limits.
Further observational studies, like MIR
spectroscopic follow-up with the JWST, will be crucial to fully understand 
 and to confirm the dust evolution triggered by 
high-mass progenitors in stellar mergers.


\begin{acknowledgements}
    MAGM, NB, GGM, HT, MW, and GK acknowledge to be funded by the European Union (ERC, CET-3PO, 101042610). Views and opinions expressed are however those of the author(s) only and do not necessarily reflect those of the European Union or the European Research Council Executive Agency. Neither the European Union nor the granting authority can be held responsible for them. We also acknowledge financial support from grant CEX2024-001451-M funded by MICIU/AEI/10.13039/501100011033
     and the support from the State Research Agency (AEI) of the Ministry of Science, Innovation and
    Universities (MICIU) of the Government of Spain under grant PID2024-155585NA-I00 funded by
     MICIU/AEI/10.13039/501100011033. TK was supported by grant SONATA-BIS no. 2018/30/E/ST9/00398 from the Polish National Science Center.

    This research has made use of the Keck Observatory Archive (KOA), which is operated by the W. M. Keck Observatory and the NASA Exoplanet Science Institute (NExScI), under contract with the National Aeronautics and Space Administration.
    This research has made use of the NASA/IPAC Infrared Science Archive, which is funded by the National Aeronautics and Space Administration and operated by the California Institute of Technology.
    This publication also makes use of data products from NEOWISE, which is a project of the Jet Propulsion Laboratory/California Institute of Technology, funded by the Planetary Science Division of the National Aeronautics and Space Administration.
    This work is based [in part] on observations made with the Spitzer Space Telescope, which was operated by the Jet Propulsion Laboratory, California Institute of Technology under a contract with NASA.
    This work presents results from the European Space Agency (ESA) space mission Gaia. Gaia data are being processed by the Gaia Data Processing and Analysis Consortium (DPAC). Funding for the DPAC is provided by national institutions, in particular the institutions participating in the Gaia MultiLateral Agreement (MLA). The Gaia mission website is \url{https://www.cosmos.esa.int/gaia}. The Gaia archive website is \url{https://archives.esac.esa.int/gaia}.
    This research also made use of Photutils, an Astropy package for detection and photometry of astronomical sources \citep{larry_bradley_2025_14889440}, APLpy, an open-source plotting package for Python
    \citep{aplpy2012,aplpy2019}, and \texttt{PypeIt} version
1.17.3,\footnote{\url{https://pypeit.readthedocs.io/en/stable/}} a Python
package for semi-automated reduction of astronomical slit-based spectroscopy
\citep{pypeit:joss_pub, pypeit:zenodo}.
\end{acknowledgements}

\bibliographystyle{aa}
\bibliography{references}

@inproceedings{1976IAUS...73...75P,
  author    = {{Paczynski}, B.},
  title     = {{Common Envelope Binaries}},
  booktitle = {Structure and Evolution of Close Binary Systems},
  year      = 1976,
  editor    = {{Eggleton}, Peter and {Mitton}, Simon and {Whelan}, John},
  series    = {IAU Symposium},
  volume    = {73},
  month     = jan,
  pages     = {75},
  adsurl    = {https://ui.adsabs.harvard.edu/abs/1976IAUS...73...75P}
}

@ARTICLE{KaminskiBlagorodnova2026,
       author = {{Kami{\'n}ski}, Tomasz and {Blagorodnova}, Nadejda},
        title = "{Red novae, their progenitors, and remnants}",
      journal = {arXiv e-prints},
         year = 2026,
        month = may,
          eid = {arXiv:2605.17005},
        pages = {arXiv:2605.17005},
          doi = {10.48550/arXiv.2605.17005},
archivePrefix = {arXiv},
       eprint = {2605.17005},
 primaryClass = {astro-ph.SR},
       adsurl = {https://ui.adsabs.harvard.edu/abs/2026arXiv260517005K}
}

@article{2017ApJ...843L..30I,
  author        = {{Ivanova}, Natalia and {da Rocha}, Cassio A. and {Van}, Kenny X. and {Nandez}, Jose L.~A.},
  title         = {{Formation of Black Hole X-Ray Binaries with Non-degenerate Donors in Globular Clusters}},
  journal       = {\apjl},
  year          = 2017,
  month         = jul,
  volume        = {843},
  number        = {2},
  eid           = {L30},
  pages         = {L30},
  doi           = {10.3847/2041-8213/aa7b76},
  archiveprefix = {arXiv},
  eprint        = {1706.07577},
  primaryclass  = {astro-ph.HE},
  adsurl        = {https://ui.adsabs.harvard.edu/abs/2017ApJ...843L..30I}
}

@article{2018A&A...617A.129K,
  author        = {{Kami{\'n}ski}, T. and {Steffen}, W. and {Tylenda}, R. and {Young}, K.~H. and {Patel}, N.~A. and {Menten}, K.~M.},
  title         = {{Submillimeter-wave emission of three Galactic red novae: cool molecular outflows produced by stellar mergers}},
  journal       = {\aap},
  year          = 2018,
  month         = oct,
  volume        = {617},
  eid           = {A129},
  pages         = {A129},
  doi           = {10.1051/0004-6361/201833165},
  archiveprefix = {arXiv},
  eprint        = {1804.01610},
  primaryclass  = {astro-ph.SR},
  adsurl        = {https://ui.adsabs.harvard.edu/abs/2018A&A...617A.129K}
}

@article{2020MNRAS.496.5503B,
  author        = {{Blagorodnova}, N. and {Karambelkar}, V. and {Adams}, S.~M. and {Kasliwal}, M.~M. and {Kochanek}, C.~S. and {Dong}, S. and {Campbell}, H. and {Hodgkin}, S. and {Jencson}, J.~E. and {Johansson}, J. and {Koz{\l}owski}, S. and {Laher}, R.~R. and {Masci}, F. and {Nugent}, P. and {Rebbapragada}, U.},
  title         = {{Progenitor, precursor, and evolution of the dusty remnant of the stellar merger M31-LRN-2015}},
  journal       = {\mnras},
  year          = 2020,
  month         = aug,
  volume        = {496},
  number        = {4},
  pages         = {5503-5517},
  doi           = {10.1093/mnras/staa1872},
  archiveprefix = {arXiv},
  eprint        = {2004.04757},
  primaryclass  = {astro-ph.SR},
  adsurl        = {https://ui.adsabs.harvard.edu/abs/2020MNRAS.496.5503B}
}

@article{2024MNRAS.533..464B,
  author        = {{Berm{\'u}dez-Bustamante}, Luis C. and {De Marco}, Orsola and {Siess}, Lionel and {Price}, Daniel J. and {Gonz{\'a}lez-Bol{\'\i}var}, Miguel and {Lau}, Mike Y.~M. and {Mu}, Chunliang and {Hirai}, Ryosuke and {Danilovich}, Ta{\"\i}ssa and {Kasliwal}, Mansi M.},
  title         = {{Dust formation in common envelope binary interactions - II: 3D simulations with self-consistent dust formation}},
  journal       = {\mnras},
  year          = 2024,
  month         = sep,
  volume        = {533},
  number        = {1},
  pages         = {464-481},
  doi           = {10.1093/mnras/stae1841},
  archiveprefix = {arXiv},
  eprint        = {2401.03644},
  primaryclass  = {astro-ph.SR},
  adsurl        = {https://ui.adsabs.harvard.edu/abs/2024MNRAS.533..464B}
}

@article{2011ApJ...733..124S,
  author        = {{Shappee}, Benjamin J. and {Stanek}, K.~Z.},
  title         = {{A New Cepheid Distance to the Giant Spiral M101 Based on Image Subtraction of Hubble Space Telescope/Advanced Camera for Surveys Observations}},
  journal       = {\apj},
  year          = 2011,
  month         = jun,
  volume        = {733},
  number        = {2},
  eid           = {124},
  pages         = {124},
  doi           = {10.1088/0004-637X/733/2/124},
  archiveprefix = {arXiv},
  eprint        = {1012.3747},
  primaryclass  = {astro-ph.CO},
  adsurl        = {https://ui.adsabs.harvard.edu/abs/2011ApJ...733..124S}
}

@article{2011ApJ...737..103S,
  author        = {{Schlafly}, Edward F. and {Finkbeiner}, Douglas P.},
  title         = {{Measuring Reddening with Sloan Digital Sky Survey Stellar Spectra and Recalibrating SFD}},
  journal       = {\apj},
  year          = 2011,
  month         = aug,
  volume        = {737},
  number        = {2},
  eid           = {103},
  pages         = {103},
  doi           = {10.1088/0004-637X/737/2/103},
  archiveprefix = {arXiv},
  eprint        = {1012.4804},
  primaryclass  = {astro-ph.GA},
  adsurl        = {https://ui.adsabs.harvard.edu/abs/2011ApJ...737..103S}
}

@article{2017ApJ...834..107B,
  author        = {{Blagorodnova}, N. and {Kotak}, R. and {Polshaw}, J. and {Kasliwal}, M.~M. and {Cao}, Y. and {Cody}, A.~M. and {Doran}, G.~B. and {Elias-Rosa}, N. and {Fraser}, M. and {Fremling}, C. and {Gonzalez-Fernandez}, C. and {Harmanen}, J. and {Jencson}, J. and {Kankare}, E. and {Kudritzki}, R.-P. and {Kulkarni}, S.~R. and {Magnier}, E. and {Manulis}, I. and {Masci}, F.~J. and {Mattila}, S. and {Nugent}, P. and {Ochner}, P. and {Pastorello}, A. and {Reynolds}, T. and {Smith}, K. and {Sollerman}, J. and {Taddia}, F. and {Terreran}, G. and {Tomasella}, L. and {Turatto}, M. and {Vreeswijk}, P.~M. and {Wozniak}, P. and {Zaggia}, S.},
  title         = {{Common Envelope Ejection for a Luminous Red Nova in M101}},
  journal       = {\apj},
  year          = 2017,
  month         = jan,
  volume        = {834},
  number        = {2},
  eid           = {107},
  pages         = {107},
  doi           = {10.3847/1538-4357/834/2/107},
  archiveprefix = {arXiv},
  eprint        = {1607.08248},
  primaryclass  = {astro-ph.SR},
  adsurl        = {https://ui.adsabs.harvard.edu/abs/2017ApJ...834..107B}
}

@software{larry_bradley_2025_14889440,
  author    = {Larry Bradley and
               Brigitta Sip{\H o}cz and
               Thomas Robitaille and
               Erik Tollerud and
               Z\`e Vin{\'{\i}}cius and
               Christoph Deil and
               Kyle Barbary and
               Tom J Wilson and
               Ivo Busko and
               Axel Donath and
               Hans Moritz G{\"u}nther and
               Mihai Cara and
               P. L. Lim and
               Sebastian Me{\ss}linger and
               Zach Burnett and
               Simon Conseil and
               Michael Droettboom and
               Azalee Bostroem and
               E. M. Bray and
               Lars Andersen Bratholm and
               William Jamieson and
               Adam Ginsburg and
               Geert Barentsen and
               Matt Craig and
               Sergio Pascual and
               Shivangee Rathi and
               Marshall Perrin and
               Brett M. Morris},
  title     = {astropy/photutils: 2.2.0},
  month     = feb,
  year      = 2025,
  publisher = {Zenodo},
  version   = {2.2.0},
  doi       = {10.5281/zenodo.14889440},
  url       = {https://doi.org/10.5281/zenodo.14889440},
  swhid     = {swh:1:dir:11159107f27a28985192ed1118b1f2055709d093
               ;origin=https://doi.org/10.5281/zenodo.596036;visi
               t=swh:1:snp:ae8c4a55d349d43e53cfe9ce92e678fcfe840f
               3b;anchor=swh:1:rel:0117f67e8888adcdfc85308287dd9c
               854b466389;path=astropy-photutils-ffb96c5
               }
}

@article{2004ApJS..154...10F,
  author        = {{Fazio}, G.~G. and {Hora}, J.~L. and {Allen}, L.~E. and {Ashby}, M.~L.~N. and {Barmby}, P. and {Deutsch}, L.~K. and {Huang}, J.-S. and {Kleiner}, S. and {Marengo}, M. and {Megeath}, S.~T. and {Melnick}, G.~J. and {Pahre}, M.~A. and {Patten}, B.~M. and {Polizotti}, J. and {Smith}, H.~A. and {Taylor}, R.~S. and {Wang}, Z. and {Willner}, S.~P. and {Hoffmann}, W.~F. and {Pipher}, J.~L. and {Forrest}, W.~J. and {McMurty}, C.~W. and {McCreight}, C.~R. and {McKelvey}, M.~E. and {McMurray}, R.~E. and {Koch}, D.~G. and {Moseley}, S.~H. and {Arendt}, R.~G. and {Mentzell}, J.~E. and {Marx}, C.~T. and {Losch}, P. and {Mayman}, P. and {Eichhorn}, W. and {Krebs}, D. and {Jhabvala}, M. and {Gezari}, D.~Y. and {Fixsen}, D.~J. and {Flores}, J. and {Shakoorzadeh}, K. and {Jungo}, R. and {Hakun}, C. and {Workman}, L. and {Karpati}, G. and {Kichak}, R. and {Whitley}, R. and {Mann}, S. and {Tollestrup}, E.~V. and {Eisenhardt}, P. and {Stern}, D. and {Gorjian}, V. and {Bhattacharya}, B. and {Carey}, S. and {Nelson}, B.~O. and {Glaccum}, W.~J. and {Lacy}, M. and {Lowrance}, P.~J. and {Laine}, S. and {Reach}, W.~T. and {Stauffer}, J.~A. and {Surace}, J.~A. and {Wilson}, G. and {Wright}, E.~L. and {Hoffman}, A. and {Domingo}, G. and {Cohen}, M.},
  title         = {{The Infrared Array Camera (IRAC) for the Spitzer Space Telescope}},
  journal       = {\apjs},
  year          = 2004,
  month         = sep,
  volume        = {154},
  number        = {1},
  pages         = {10-17},
  doi           = {10.1086/422843},
  archiveprefix = {arXiv},
  eprint        = {astro-ph/0405616},
  primaryclass  = {astro-ph},
  adsurl        = {https://ui.adsabs.harvard.edu/abs/2004ApJS..154...10F}
}

@article{2017ApJ...839...88K,
  author        = {{Kasliwal}, Mansi M. and {Bally}, John and {Masci}, Frank and {Cody}, Ann Marie and {Bond}, Howard E. and {Jencson}, Jacob E. and {Tinyanont}, Samaporn and {Cao}, Yi and {Contreras}, Carlos and {Dykhoff}, Devin A. and {Amodeo}, Samuel and {Armus}, Lee and {Boyer}, Martha and {Cantiello}, Matteo and {Carlon}, Robert L. and {Cass}, Alexander C. and {Cook}, David and {Corgan}, David T. and {Faella}, Joseph and {Fox}, Ori D. and {Green}, Wayne and {Gehrz}, R.~D. and {Helou}, George and {Hsiao}, Eric and {Johansson}, Joel and {Khan}, Rubab M. and {Lau}, Ryan M. and {Langer}, Norbert and {Levesque}, Emily and {Milne}, Peter and {Mohamed}, Shazrene and {Morrell}, Nidia and {Monson}, Andy and {Moore}, Anna and {Ofek}, Eran O. and {O' Sullivan}, Donal and {Parthasarathy}, Mudumba and {Perez}, Andres and {Perley}, Daniel A. and {Phillips}, Mark and {Prince}, Thomas A. and {Shenoy}, Dinesh and {Smith}, Nathan and {Surace}, Jason and {Van Dyk}, Schuyler D. and {Whitelock}, Patricia A. and {Williams}, Robert},
  title         = {{SPIRITS: Uncovering Unusual Infrared Transients with Spitzer}},
  journal       = {\apj},
  year          = 2017,
  month         = apr,
  volume        = {839},
  number        = {2},
  eid           = {88},
  pages         = {88},
  doi           = {10.3847/1538-4357/aa6978},
  archiveprefix = {arXiv},
  eprint        = {1701.01151},
  primaryclass  = {astro-ph.HE},
  adsurl        = {https://ui.adsabs.harvard.edu/abs/2017ApJ...839...88K}
}

@article{2000PASP..112..315W,
  author   = {{Wizinowich}, P. and {Acton}, D.~S. and {Shelton}, C. and {Stomski}, P. and {Gathright}, J. and {Ho}, K. and {Lupton}, W. and {Tsubota}, K. and {Lai}, O. and {Max}, C. and et al.},
  title    = {{First Light Adaptive Optics Images from the Keck II Telescope: A New Era of High Angular Resolution Imagery}},
  journal  = {\pasp},
  year     = 2000,
  month    = mar,
  volume   = {112},
  number   = {769},
  pages    = {315-319},
  doi      = {10.1086/316543},
  adsurl   = {https://ui.adsabs.harvard.edu/abs/2000PASP..112..315W}
}

@article{2012Sci...337..444S,
  author        = {{Sana}, H. and {de Mink}, S.~E. and {de Koter}, A. and {Langer}, N. and {Evans}, C.~J. and {Gieles}, M. and {Gosset}, E. and {Izzard}, R.~G. and {Le Bouquin}, J.-B. and {Schneider}, F.~R.~N.},
  title         = {{Binary Interaction Dominates the Evolution of Massive Stars}},
  journal       = {Science},
  year          = 2012,
  month         = jul,
  volume        = {337},
  number        = {6093},
  pages         = {444},
  doi           = {10.1126/science.1223344},
  archiveprefix = {arXiv},
  eprint        = {1207.6397},
  primaryclass  = {astro-ph.SR},
  adsurl        = {https://ui.adsabs.harvard.edu/abs/2012Sci...337..444S}
}

@article{2025ApJ...994L..41K,
  author        = {{Kirilov}, Anthony and {Calder{\'o}n}, Diego and {Pejcha}, Ond{\v{r}}ej and {Duffell}, Paul C.},
  title         = {{Two-dimensional Radiation-hydrodynamic Simulations of Luminous Red Novae}},
  journal       = {\apjl},
  year          = 2025,
  month         = dec,
  volume        = {994},
  number        = {2},
  eid           = {L41},
  pages         = {L41},
  doi           = {10.3847/2041-8213/ae1ae7},
  archiveprefix = {arXiv},
  eprint        = {2508.09257},
  primaryclass  = {astro-ph.SR},
  adsurl        = {https://ui.adsabs.harvard.edu/abs/2025ApJ...994L..41K}
}

@article{2017ApJ...846..170T,
  author        = {{Tauris}, T.~M. and {Kramer}, M. and {Freire}, P.~C.~C. and {Wex}, N. and {Janka}, H.-T. and {Langer}, N. and {Podsiadlowski}, Ph. and {Bozzo}, E. and {Chaty}, S. and {Kruckow}, M.~U. and {van den Heuvel}, E.~P.~J. and {Antoniadis}, J. and {Breton}, R.~P. and {Champion}, D.~J.},
  title         = {{Formation of Double Neutron Star Systems}},
  journal       = {\apj},
  year          = 2017,
  month         = sep,
  volume        = {846},
  number        = {2},
  eid           = {170},
  pages         = {170},
  doi           = {10.3847/1538-4357/aa7e89},
  archiveprefix = {arXiv},
  eprint        = {1706.09438},
  primaryclass  = {astro-ph.HE},
  adsurl        = {https://ui.adsabs.harvard.edu/abs/2017ApJ...846..170T}
}

@article{2018A&A...615A..91B,
  author        = {{Belczynski}, K. and {Askar}, A. and {Arca-Sedda}, M. and {Chruslinska}, M. and {Donnari}, M. and {Giersz}, M. and {Benacquista}, M. and {Spurzem}, R. and {Jin}, D. and {Wiktorowicz}, G. and {Belloni}, D.},
  title         = {{The origin of the first neutron star - neutron star merger}},
  journal       = {\aap},
  year          = 2018,
  month         = jul,
  volume        = {615},
  eid           = {A91},
  pages         = {A91},
  doi           = {10.1051/0004-6361/201732428},
  archiveprefix = {arXiv},
  eprint        = {1712.00632},
  primaryclass  = {astro-ph.HE},
  adsurl        = {https://ui.adsabs.harvard.edu/abs/2018A&A...615A..91B}
}

@article{2000A&A...360.1011N,
  author        = {{Nelemans}, G. and {Verbunt}, F. and {Yungelson}, L.~R. and {Portegies Zwart}, Simon F.},
  title         = {{Reconstructing the evolution of double helium white dwarfs: envelope loss without spiral-in}},
  journal       = {\aap},
  year          = 2000,
  month         = aug,
  volume        = {360},
  pages         = {1011-1018},
  doi           = {10.48550/arXiv.astro-ph/0006216},
  archiveprefix = {arXiv},
  eprint        = {astro-ph/0006216},
  primaryclass  = {astro-ph},
  adsurl        = {https://ui.adsabs.harvard.edu/abs/2000A&A...360.1011N}
}

@article{2018MNRAS.481.4009V,
  author        = {{Vigna-G{\'o}mez}, Alejandro and {Neijssel}, Coenraad J. and {Stevenson}, Simon and {Barrett}, Jim W. and {Belczynski}, Krzysztof and {Justham}, Stephen and {de Mink}, Selma E. and {M{\"u}ller}, Bernhard and {Podsiadlowski}, Philipp and {Renzo}, Mathieu and {Sz{\'e}csi}, Dorottya and {Mandel}, Ilya},
  title         = {{On the formation history of Galactic double neutron stars}},
  journal       = {\mnras},
  year          = 2018,
  month         = dec,
  volume        = {481},
  number        = {3},
  pages         = {4009-4029},
  doi           = {10.1093/mnras/sty2463},
  archiveprefix = {arXiv},
  eprint        = {1805.07974},
  primaryclass  = {astro-ph.SR},
  adsurl        = {https://ui.adsabs.harvard.edu/abs/2018MNRAS.481.4009V}
}

@article{2024MNRAS.527.9145G,
  author        = {{Gonz{\'a}lez-Bol{\'\i}var}, Miguel and {De Marco}, Orsola and {Berm{\'u}dez-Bustamante}, Luis C. and {Siess}, Lionel and {Price}, Daniel J.},
  title         = {{Dust formation in common envelope binary interaction - I: 3D simulations using the Bowen approximation}},
  journal       = {\mnras},
  year          = 2024,
  month         = jan,
  volume        = {527},
  number        = {3},
  pages         = {9145-9158},
  doi           = {10.1093/mnras/stad3748},
  archiveprefix = {arXiv},
  eprint        = {2306.16609},
  primaryclass  = {astro-ph.SR},
  adsurl        = {https://ui.adsabs.harvard.edu/abs/2024MNRAS.527.9145G}
}

@article{2016A&A...590A..65M,
  author        = {{Micelotta}, Elisabetta R. and {Dwek}, Eli and {Slavin}, Jonathan D.},
  title         = {{Dust destruction by the reverse shock in the Cassiopeia A supernova remnant}},
  journal       = {\aap},
  year          = 2016,
  month         = may,
  volume        = {590},
  eid           = {A65},
  pages         = {A65},
  doi           = {10.1051/0004-6361/201527350},
  archiveprefix = {arXiv},
  eprint        = {1602.02754},
  primaryclass  = {astro-ph.GA},
  adsurl        = {https://ui.adsabs.harvard.edu/abs/2016A&A...590A..65M}
}

@ARTICLE{2015ApJ...814..109B,
       author = {{Banerjee}, Dipankar. P.~K. and {Nuth}, III, Joseph A. and {Misselt}, Karl A. and {Varricatt}, Watson P. and {Sand}, David and {Ashok}, N.~M. and {Su}, K.~Y.~L. and {Marion}, G.~H. and {Marengo}, Massimo},
        title = "{Evolution of the Dust in V4332 Sagittarii}",
      journal = {\apj},
         year = 2015,
        month = dec,
       volume = {814},
       number = {2},
          eid = {109},
        pages = {109},
          doi = {10.1088/0004-637X/814/2/109},
archivePrefix = {arXiv},
       eprint = {1510.05074},
 primaryClass = {astro-ph.SR},
       adsurl = {https://ui.adsabs.harvard.edu/abs/2015ApJ...814..109B}
}

@ARTICLE{2022A&A...667A...4C,
       author = {{Cai}, Y.-Z. and {Pastorello}, A. and {Fraser}, M. and {Wang}, X.-F. and {Filippenko}, A.~V. and {Reguitti}, A. and {Patra}, K.~C. and {Goranskij}, V.~P. and {Barsukova}, E.~A. and {Brink}, T.~G. and {Elias-Rosa}, N. and {Stevance}, H.~F. and {Zheng}, W. and {Yang}, Y. and {Atapin}, K.~E. and {Benetti}, S. and {de Boer}, T.~J.~L. and {Bose}, S. and {Burke}, J. and {Byrne}, R. and {Cappellaro}, E. and {Chambers}, K.~C. and {Chen}, W.-L. and {Emami}, N. and {Gao}, H. and {Hiramatsu}, D. and {Howell}, D.~A. and {Huber}, M.~E. and {Kankare}, E. and {Kelly}, P.~L. and {Kotak}, R. and {Kravtsov}, T. and {Lander}, V. Yu. and {Li}, Z.-T. and {Lin}, C.-C. and {Lundqvist}, P. and {Magnier}, E.~A. and {Malygin}, E.~A. and {Maslennikova}, N.~A. and {Matilainen}, K. and {Mazzali}, P.~A. and {McCully}, C. and {Mo}, J. and {Moran}, S. and {Newsome}, M. and {Oparin}, D.~V. and {Padilla Gonzalez}, E. and {Reynolds}, T.~M. and {Shatsky}, N.~I. and {Smartt}, S.~J. and {Smith}, K.~W. and {Stritzinger}, M.~D. and {Tatarnikov}, A.~M. and {Terreran}, G. and {Uklein}, R.~I. and {Valerin}, G. and {Vallely}, P.~J. and {Vozyakova}, O.~V. and {Wainscoat}, R. and {Yan}, S.-Y. and {Zhang}, J.-J. and {Zhang}, T.-M. and {Zheltoukhov}, S.~G. and {Dastidar}, R. and {Fulton}, M. and {Galbany}, L. and {Gangopadhyay}, A. and {Ge}, H.-W. and {Guti{\'e}rrez}, C.~P. and {Lin}, H. and {Misra}, K. and {Ou}, Z.-W. and {Salmaso}, I. and {Tartaglia}, L. and {Xiao}, L. and {Zhang}, X.-H.},
        title = "{Forbidden hugs in pandemic times. III. Observations of the luminous red nova AT 2021biy in the nearby galaxy NGC 4631}",
      journal = {\aap},
         year = 2022,
        month = nov,
       volume = {667},
          eid = {A4},
        pages = {A4},
          doi = {10.1051/0004-6361/202244393},
archivePrefix = {arXiv},
       eprint = {2207.00734},
 primaryClass = {astro-ph.SR},
       adsurl = {https://ui.adsabs.harvard.edu/abs/2022A&A...667A...4C}
}

@ARTICLE{2023A&A...671A.158P,
       author = {{Pastorello}, A. and {Valerin}, G. and {Fraser}, M. and {Reguitti}, A. and {Elias-Rosa}, N. and {Filippenko}, A.~V. and {Rojas-Bravo}, C. and {Tartaglia}, L. and {Reynolds}, T.~M. and {Valenti}, S. and {Andrews}, J.~E. and {Ashall}, C. and {Bostroem}, K.~A. and {Brink}, T.~G. and {Burke}, J. and {Cai}, Y.-Z. and {Cappellaro}, E. and {Coulter}, D.~A. and {Dastidar}, R. and {Davis}, K.~W. and {Dimitriadis}, G. and {Fiore}, A. and {Foley}, R.~J. and {Fugazza}, D. and {Galbany}, L. and {Gangopadhyay}, A. and {Geier}, S. and {Guti{\'e}rrez}, C.~P. and {Haislip}, J. and {Hiramatsu}, D. and {Holmbo}, S. and {Howell}, D.~A. and {Hsiao}, E.~Y. and {Hung}, T. and {Jha}, S.~W. and {Kankare}, E. and {Karamehmetoglu}, E. and {Kilpatrick}, C.~D. and {Kotak}, R. and {Kouprianov}, V. and {Kravtsov}, T. and {Kumar}, S. and {Li}, Z.-T. and {Lundquist}, M.~J. and {Lundqvist}, P. and {Matilainen}, K. and {Mazzali}, P.~A. and {McCully}, C. and {Misra}, K. and {Morales-Garoffolo}, A. and {Moran}, S. and {Morrell}, N. and {Newsome}, M. and {Padilla Gonzalez}, E. and {Pan}, Y.-C. and {Pellegrino}, C. and {Phillips}, M.~M. and {Pignata}, G. and {Piro}, A.~L. and {Reichart}, D.~E. and {Rest}, A. and {Salmaso}, I. and {Sand}, D.~J. and {Siebert}, M.~R. and {Smartt}, S.~J. and {Smith}, K.~W. and {Srivastav}, S. and {Stritzinger}, M.~D. and {Taggart}, K. and {Tinyanont}, S. and {Yan}, S.-Y. and {Wang}, L. and {Wang}, X.-F. and {Williams}, S.~C. and {Wyatt}, S. and {Zhang}, T.-M. and {de Boer}, T. and {Chambers}, K. and {Gao}, H. and {Magnier}, E.},
        title = "{Forbidden hugs in pandemic times. IV. Panchromatic evolution of three luminous red novae}",
      journal = {\aap},
         year = 2023,
        month = mar,
       volume = {671},
          eid = {A158},
        pages = {A158},
          doi = {10.1051/0004-6361/202244684},
archivePrefix = {arXiv},
       eprint = {2208.02782},
 primaryClass = {astro-ph.SR},
       adsurl = {https://ui.adsabs.harvard.edu/abs/2023A&A...671A.158P}
}

@ARTICLE{2004ApJ...615L..53B,
       author = {{Banerjee}, Dipankar P.~K. and {Varricatt}, Watson P. and {Ashok}, Nagarhalli M.},
        title = "{L and M Band Infrared Studies of V4332 Sagittarii: Detection of the Water Ice Absorption Band at 3.05 {\ensuremath{\mu}}m and the CO Fundamental Band in Emission}",
      journal = {\apjl},
         year = 2004,
        month = nov,
       volume = {615},
       number = {1},
        pages = {L53-L56},
          doi = {10.1086/425963},
archivePrefix = {arXiv},
       eprint = {astro-ph/0409393},
 primaryClass = {astro-ph},
       adsurl = {https://ui.adsabs.harvard.edu/abs/2004ApJ...615L..53B}
}

@ARTICLE{2007ApJ...666L..25B,
       author = {{Banerjee}, D.~P.~K. and {Misselt}, K.~A. and {Su}, K.~Y.~L. and {Ashok}, N.~M. and {Smith}, P.~S.},
        title = "{Spitzer Observations of V4332 Sagittarii: Detection of Alumina Dust}",
      journal = {\apjl},
         year = 2007,
        month = sep,
       volume = {666},
       number = {1},
        pages = {L25-L28},
          doi = {10.1086/521528},
archivePrefix = {arXiv},
       eprint = {0707.3119},
 primaryClass = {astro-ph},
       adsurl = {https://ui.adsabs.harvard.edu/abs/2007ApJ...666L..25B}
}

@ARTICLE{2021AJ....162..183W,
       author = {{Woodward}, C.~E. and {Evans}, A. and {Banerjee}, D.~P.~K. and {Liimets}, T. and {Djupvik}, A.~A. and {Starrfield}, S. and {Clayton}, G.~C. and {Eyres}, S.~P.~S. and {Gehrz}, R.~D. and {Wagner}, R.~M.},
        title = "{The Infrared Evolution of Dust in V838 Monocerotis}",
      journal = {\aj},
         year = 2021,
        month = nov,
       volume = {162},
       number = {5},
          eid = {183},
        pages = {183},
          doi = {10.3847/1538-3881/ac1f1e},
archivePrefix = {arXiv},
       eprint = {2108.08149},
 primaryClass = {astro-ph.SR},
       adsurl = {https://ui.adsabs.harvard.edu/abs/2021AJ....162..183W}
}

@ARTICLE{2026ApJ...999L..35B,
       author = {{Banerjee}, D.~P.~K. and {Evans}, A. and {Varricatt}, Watson P. and {Ashok}, N.~M.},
        title = "{Evidence for the Merger Hypothesis in V4332 Sgr: A Low $^{12}$C/$^{13}$C Ratio and Multiple Outbursts}",
      journal = {\apjl},
         year = 2026,
        month = mar,
       volume = {999},
       number = {2},
          eid = {L35},
        pages = {L35},
          doi = {10.3847/2041-8213/ae46a3},
archivePrefix = {arXiv},
       eprint = {2602.14781},
 primaryClass = {astro-ph.SR},
       adsurl = {https://ui.adsabs.harvard.edu/abs/2026ApJ...999L..35B}
}

@ARTICLE{2019AJ....158..257B,
	author = {{Boone}, Kyle},
	title = "{Avocado: Photometric Classification of Astronomical Transients with Gaussian Process Augmentation}",
	journal = {\aj},
	year = 2019,
	month = dec,
	volume = {158},
	number = {6},
	eid = {257},
	pages = {257},
	doi = {10.3847/1538-3881/ab5182},
	archivePrefix = {arXiv},
	eprint = {1907.04690},
	primaryClass = {astro-ph.IM},
	adsurl = {https://ui.adsabs.harvard.edu/abs/2019AJ....158..257B}
}

@ARTICLE{2026ApJ...999...16K,
	author = {{Karambelkar}, Viraj and {Kasliwal}, Mansi M. and {Lau}, Ryan M. and {Jencson}, Jacob E. and {Blagorodnova}, Nadejda and {G{\'o}mez-Mu{\~n}oz}, Marco A. and {Tranin}, Hugo and {Wavasseur}, Maxime and {Shahbandeh}, Melissa and {De}, Kishalay},
	title = "{Hot Springs and Dust Reservoirs: JWST Reveals the Dusty, Molecular Aftermath of Extragalactic Stellar Mergers}",
	journal = {\apj},
	year = 2026,
	month = mar,
	volume = {999},
	number = {1},
	eid = {16},
	pages = {16},
	doi = {10.3847/1538-4357/ae38bf},
	archivePrefix = {arXiv},
	eprint = {2508.03932},
	primaryClass = {astro-ph.SR},
	adsurl = {https://ui.adsabs.harvard.edu/abs/2026ApJ...999...16K}
}

@ARTICLE{1997MNRAS.287..799I,
	author = {{Ivezic}, Zeljko and {Elitzur}, Moshe},
	title = "{Self-similarity and scaling behaviour of infrared emission from radiatively heated dust - I. Theory}",
	journal = {\mnras},
	year = 1997,
	month = jun,
	volume = {287},
	number = {4},
	pages = {799-811},
	doi = {10.1093/mnras/287.4.799},
	archivePrefix = {arXiv},
	eprint = {astro-ph/9612164},
	primaryClass = {astro-ph},
	adsurl = {https://ui.adsabs.harvard.edu/abs/1997MNRAS.287..799I}
}

@software{1999ascl.soft11001I,
	author = {{Ivezic}, Zeljko and {Nenkova}, Maia and {Elitzur}, Moshe},
	title = "{DUSTY: Radiation transport in a dusty environment}",
	howpublished = {Astrophysics Source Code Library, record ascl:9911.001},
	year = 1999,
	month = nov,
	eid = {ascl:9911.001},
	archivePrefix = {ascl},
	eprint = {9911.001},
	adsurl = {https://ui.adsabs.harvard.edu/abs/1999ascl.soft11001I}
}

@INPROCEEDINGS{2000ASPC..196...77N,
	author = {{Nenkova}, M. and {Ivezi{\'c}}, {\v{Z}}. and {Elitzur}, M.},
	title = "{DUSTY: a Publicly Available Code for Modeling Dust Emission}",
	booktitle = {Thermal Emission Spectroscopy and Analysis of Dust, Disks, and Regoliths},
	year = 2000,
	editor = {{Sitko}, Michael L. and {Sprague}, Ann L. and {Lynch}, David K.},
	series = {Astronomical Society of the Pacific Conference Series},
	volume = {196},
	month = mar,
	pages = {77-82},
	adsurl = {https://ui.adsabs.harvard.edu/abs/2000ASPC..196...77N}
}

@ARTICLE{2013PASP..125..306F,
	author = {{Foreman-Mackey}, Daniel and {Hogg}, David W. and {Lang}, Dustin and {Goodman}, Jonathan},
	title = "{emcee: The MCMC Hammer}",
	journal = {\pasp},
	year = 2013,
	month = mar,
	volume = {125},
	number = {925},
	pages = {306},
	doi = {10.1086/670067},
	archivePrefix = {arXiv},
	eprint = {1202.3665},
	primaryClass = {astro-ph.IM},
	adsurl = {https://ui.adsabs.harvard.edu/abs/2013PASP..125..306F}
}

@ARTICLE{2025ApJ...983...87L,
	author = {{Lau}, Ryan M. and {Jencson}, Jacob E. and {Salyk}, Colette and {De}, Kishalay and {Fox}, Ori D. and {Hankins}, Matthew J. and {Kasliwal}, Mansi M. and {Keyes}, Charles D. and {Macleod}, Morgan and {Ressler}, Michael E. and {Rose}, Sam},
	title = "{Revealing a Main-sequence Star that Consumed a Planet with JWST}",
	journal = {\apj},
	year = 2025,
	month = apr,
	volume = {983},
	number = {2},
	eid = {87},
	pages = {87},
	doi = {10.3847/1538-4357/adb429},
	archivePrefix = {arXiv},
	eprint = {2504.07275},
	primaryClass = {astro-ph.SR},
	adsurl = {https://ui.adsabs.harvard.edu/abs/2025ApJ...983...87L}
}

@ARTICLE{1984ApJ...285...89D,
	author = {{Draine}, B.~T. and {Lee}, H.~M.},
	title = "{Optical Properties of Interstellar Graphite and Silicate Grains}",
	journal = {\apj},
	year = 1984,
	month = oct,
	volume = {285},
	pages = {89},
	doi = {10.1086/162480},
	adsurl = {https://ui.adsabs.harvard.edu/abs/1984ApJ...285...89D}
}

@ARTICLE{1977ApJ...217..425M,
	author = {{Mathis}, J.~S. and {Rumpl}, W. and {Nordsieck}, K.~H.},
	title = "{The size distribution of interstellar grains.}",
	journal = {\apj},
	year = 1977,
	month = oct,
	volume = {217},
	pages = {425-433},
	doi = {10.1086/155591},
	adsurl = {https://ui.adsabs.harvard.edu/abs/1977ApJ...217..425M}
}

@ARTICLE{2005A&A...436.1009T,
       author = {{Tylenda}, R.},
        title = "{Evolution of V838 Monocerotis during and after the 2002 eruption}",
      journal = {\aap},
         year = 2005,
        month = jun,
       volume = {436},
       number = {3},
        pages = {1009-1020},
          doi = {10.1051/0004-6361:20052800},
archivePrefix = {arXiv},
       eprint = {astro-ph/0502060},
 primaryClass = {astro-ph},
       adsurl = {https://ui.adsabs.harvard.edu/abs/2005A&A...436.1009T}
}

@ARTICLE{2021A&A...653A.134B,
       author = {{Blagorodnova}, Nadejda and {Klencki}, Jakub and {Pejcha}, Ond{\v{r}}ej and {Vreeswijk}, Paul M. and {Bond}, Howard E. and {Burdge}, Kevin B. and {De}, Kishalay and {Fremling}, Christoffer and {Gehrz}, Robert D. and {Jencson}, Jacob E. and {Kasliwal}, Mansi M. and {Kupfer}, Thomas and {Lau}, Ryan M. and {Masci}, Frank J. and {Rich}, Michael R.},
        title = "{The luminous red nova AT 2018bwo in NGC 45 and its binary yellow supergiant progenitor}",
      journal = {\aap},
         year = 2021,
        month = sep,
       volume = {653},
          eid = {A134},
        pages = {A134},
          doi = {10.1051/0004-6361/202140525},
archivePrefix = {arXiv},
       eprint = {2102.05662},
 primaryClass = {astro-ph.SR},
       adsurl = {https://ui.adsabs.harvard.edu/abs/2021A&A...653A.134B}
}

@ARTICLE{2017MNRAS.471.3200M,
       author = {{Metzger}, Brian D. and {Pejcha}, Ond{\v{r}}ej},
        title = "{Shock-powered light curves of luminous red novae as signatures of pre-dynamical mass-loss in stellar mergers}",
      journal = {\mnras},
         year = 2017,
        month = nov,
       volume = {471},
       number = {3},
        pages = {3200-3211},
          doi = {10.1093/mnras/stx1768},
archivePrefix = {arXiv},
       eprint = {1705.03895},
 primaryClass = {astro-ph.HE},
       adsurl = {https://ui.adsabs.harvard.edu/abs/2017MNRAS.471.3200M}
}

@ARTICLE{2003PASP..115..362R,
       author = {{Rayner}, J.~T. and {Toomey}, D.~W. and {Onaka}, P.~M. and {Denault}, A.~J. and {Stahlberger}, W.~E. and {Vacca}, W.~D. and {Cushing}, M.~C. and {Wang}, S.},
        title = "{SpeX: A Medium-Resolution 0.8-5.5 Micron Spectrograph and Imager for the NASA Infrared Telescope Facility}",
      journal = {\pasp},
         year = 2003,
        month = mar,
       volume = {115},
       number = {805},
        pages = {362-382},
          doi = {10.1086/367745},
       adsurl = {https://ui.adsabs.harvard.edu/abs/2003PASP..115..362R}
}

@ARTICLE{2023ApJ...948..137K,
       author = {{Karambelkar}, Viraj R. and {Kasliwal}, Mansi M. and {Blagorodnova}, Nadejda and {Sollerman}, Jesper and {Aloisi}, Robert and {Anand}, Shreya G. and {Andreoni}, Igor and {Brink}, Thomas G. and {Bruch}, Rachel and {Cook}, David and et al.},
        title = "{Volumetric Rates of Luminous Red Novae and Intermediate-luminosity Red Transients with the Zwicky Transient Facility}",
      journal = {\apj},
         year = 2023,
        month = may,
       volume = {948},
       number = {2},
          eid = {137},
        pages = {137},
          doi = {10.3847/1538-4357/acc2b9},
archivePrefix = {arXiv},
       eprint = {2211.05141},
 primaryClass = {astro-ph.HE},
       adsurl = {https://ui.adsabs.harvard.edu/abs/2023ApJ...948..137K}
}

@ARTICLE{2004ApJ...607..460L,
       author = {{Lynch}, David K. and {Rudy}, Richard J. and {Russell}, Ray W. and {Mazuk}, S. and {Venturini}, Catherine C. and {Dimpfl}, W. and {Bernstein}, Lawrence S. and {Sitko}, Michael L. and {Fajardo-Acosta}, Sergio and {Tokunaga}, Alan and et al.},
        title = "{0.8-13 Micron Spectroscopy of V838 Monocerotis and a Model for Its Emission}",
      journal = {\apj},
         year = 2004,
        month = may,
       volume = {607},
       number = {1},
        pages = {460-473},
          doi = {10.1086/382667},
       adsurl = {https://ui.adsabs.harvard.edu/abs/2004ApJ...607..460L}
}

@ARTICLE{2015ApJS..216...15L,
       author = {{Li}, Gang and {Gordon}, Iouli E. and {Rothman}, Laurence S. and {Tan}, Yan and {Hu}, Shui-Ming and {Kassi}, Samir and {Campargue}, Alain and {Medvedev}, Emile S.},
        title = "{Rovibrational Line Lists for Nine Isotopologues of the CO Molecule in the X $^{1}${\ensuremath{\Sigma}}$^{+}$ Ground Electronic State}",
      journal = {\apjs},
         year = 2015,
        month = jan,
       volume = {216},
       number = {1},
          eid = {15},
        pages = {15},
          doi = {10.1088/0067-0049/216/1/15},
       adsurl = {https://ui.adsabs.harvard.edu/abs/2015ApJS..216...15L}
}

@ARTICLE{2005ApJ...627L.141B,
       author = {{Banerjee}, D.~P.~K. and {Barber}, R.~J. and {Ashok}, N.~M. and {Tennyson}, J.},
        title = "{Near-Infrared Water Lines in V838 Monocerotis}",
      journal = {\apjl},
         year = 2005,
        month = jul,
       volume = {627},
       number = {2},
        pages = {L141-L144},
          doi = {10.1086/432442},
archivePrefix = {arXiv},
       eprint = {astro-ph/0506403},
 primaryClass = {astro-ph},
       adsurl = {https://ui.adsabs.harvard.edu/abs/2005ApJ...627L.141B}
}

@ARTICLE{2026A&A...706A.154R,
       author = {{Reguitti}, A. and {Pastorello}, A. and {Valerin}, G.},
        title = "{The fate of the progenitors of luminous red novae: Infrared detection of LRNe years after the outburst}",
      journal = {\aap},
         year = 2026,
        month = feb,
       volume = {706},
          eid = {A154},
        pages = {A154},
          doi = {10.1051/0004-6361/202555226},
archivePrefix = {arXiv},
       eprint = {2504.14592},
 primaryClass = {astro-ph.SR},
       adsurl = {https://ui.adsabs.harvard.edu/abs/2026A&A...706A.154R}
}

@ARTICLE{Kirchschlager2022,
       author = {{Kirchschlager}, Florian and {Mattsson}, Lars and {Gent}, Frederick A.},
        title = "{Supernova induced processing of interstellar dust: impact of interstellar medium gas density and gas turbulence}",
      journal = {\mnras},
         year = 2022,
        month = jan,
       volume = {509},
       number = {3},
        pages = {3218-3234},
          doi = {10.1093/mnras/stab3059},
archivePrefix = {arXiv},
       eprint = {2109.01175},
 primaryClass = {astro-ph.GA},
       adsurl = {https://ui.adsabs.harvard.edu/abs/2022MNRAS.509.3218K}
}

@ARTICLE{Ginolfi2018,
       author = {{Ginolfi}, M. and {Graziani}, L. and {Schneider}, R. and {Marassi}, S. and {Valiante}, R. and {Dell'Agli}, F. and {Ventura}, P. and {Hunt}, L.~K.},
        title = "{Where does galactic dust come from?}",
      journal = {\mnras},
         year = 2018,
        month = feb,
       volume = {473},
       number = {4},
        pages = {4538-4543},
          doi = {10.1093/mnras/stx2572},
archivePrefix = {arXiv},
       eprint = {1707.05328},
 primaryClass = {astro-ph.GA},
       adsurl = {https://ui.adsabs.harvard.edu/abs/2018MNRAS.473.4538G}
}

@ARTICLE{Pastorello2019,
       author = {{Pastorello}, A. and {Mason}, E. and {Taubenberger}, S. and {Fraser}, M. and {Cortini}, G. and {Tomasella}, L. and {Botticella}, M.~T. and {Elias-Rosa}, N. and {Kotak}, R. and {Smartt}, S.~J. and {Benetti}, S. and {Cappellaro}, E. and {Turatto}, M. and {Tartaglia}, L. and {Djorgovski}, S.~G. and {Drake}, A.~J. and {Berton}, M. and {Briganti}, F. and {Brimacombe}, J. and {Bufano}, F. and {Cai}, Y.-Z. and {Chen}, S. and {Christensen}, E.~J. and {Ciabattari}, F. and {Congiu}, E. and {Dimai}, A. and {Inserra}, C. and {Kankare}, E. and {Magill}, L. and {Maguire}, K. and {Martinelli}, F. and {Morales-Garoffolo}, A. and {Ochner}, P. and {Pignata}, G. and {Reguitti}, A. and {Sollerman}, J. and {Spiro}, S. and {Terreran}, G. and {Wright}, D.~E.},
        title = "{Luminous red novae: Stellar mergers or giant eruptions?}",
      journal = {\aap},
         year = 2019,
        month = oct,
       volume = {630},
          eid = {A75},
        pages = {A75},
          doi = {10.1051/0004-6361/201935999},
archivePrefix = {arXiv},
       eprint = {1906.00812},
 primaryClass = {astro-ph.SR},
       adsurl = {https://ui.adsabs.harvard.edu/abs/2019A&A...630A..75P}
}

@ARTICLE{Goranskij2016AstBu,
       author = {{Goranskij}, V.~P. and {Barsukova}, E.~A. and {Spiridonova}, O.~I. and {Valeev}, A.~F. and {Fatkhullin}, T.~A. and {Moskvitin}, A.~S. and {Vozyakova}, O.~V. and {Cheryasov}, D.~V. and {Safonov}, B.~S. and {Zharova}, A.~V. and {Hancock}, T.},
        title = "{Photometry and spectroscopy of the luminous red nova PSNJ14021678+5426205 in the galaxy M101}",
      journal = {Astrophysical Bulletin},
         year = 2016,
        month = jan,
       volume = {71},
       number = {1},
        pages = {82-94},
          doi = {10.1134/S1990341316010090},
archivePrefix = {arXiv},
       eprint = {1611.04936},
 primaryClass = {astro-ph.SR},
       adsurl = {https://ui.adsabs.harvard.edu/abs/2016AstBu..71...82G}
}

@ARTICLE{Reguitti2026,
       author = {{Reguitti}, A. and {Pastorello}, A. and {Valerin}, G. and {Romanov}, F.~D. and {Siviero}, A. and {Cai}, Y.-Z. and {Ciroi}, S. and {Elias-Rosa}, N. and {Iijima}, T. and {Kankare}, E. and et al.},
        title = "{AT 2025abao: The fourth luminous red nova in M 31}",
      journal = {\aap},
         year = 2026,
        month = apr,
       volume = {709},
          eid = {A25},
        pages = {A25},
          doi = {10.1051/0004-6361/202659456},
archivePrefix = {arXiv},
       eprint = {2602.13678},
 primaryClass = {astro-ph.SR},
       adsurl = {https://ui.adsabs.harvard.edu/abs/2026A&A...709A..25R}
}

@ARTICLE{2021A&A...655A..32K,
       author = {{Kami{\'n}ski}, Tomek and {Tylenda}, Romuald and {Kiljan}, Aleksandra and {Schmidt}, Mirek and {Lisiecki}, Krzysztof and {Melis}, Carl and {Frankowski}, Adam and {Joshi}, Vishal and {Menten}, Karl M.},
        title = "{V838 Monocerotis as seen by ALMA: A remnant of a binary merger in a triple system}",
      journal = {\aap},
         year = 2021,
        month = nov,
       volume = {655},
          eid = {A32},
        pages = {A32},
          doi = {10.1051/0004-6361/202141526},
archivePrefix = {arXiv},
       eprint = {2106.07427},
 primaryClass = {astro-ph.SR},
       adsurl = {https://ui.adsabs.harvard.edu/abs/2021A&A...655A..32K}
}

@ARTICLE{2026MNRAS.tmp..926T,
       author = {{Tao}, Chenyi and {Abu El Kher}, Nariman and {El-Kork}, Nayla and {Yurchenko}, Sergei N. and {Tennyson}, Jonathan},
        title = "{ExoMol line lists - LXX: The CO fourth positive band system}",
      journal = {\mnras},
         year = 2026,
        month = may,
          doi = {10.1093/mnras/stag985},
       adsurl = {https://ui.adsabs.harvard.edu/abs/2026MNRAS.tmp..926T}
}

@ARTICLE{2026ApJ...998L..36Z,
       author = {{Zavala}, Jorge A. and {Faisst}, Andreas L. and {Aravena}, Manuel and {Casey}, Caitlin M. and {Kartaltepe}, Jeyhan S. and {Martinez}, III, Felix and {Silverman}, John D. and {Toft}, Sune and {Treister}, Ezequiel and {Akins}, Hollis B. and {Algera}, Hiddo and {Barboza}, Karina and {Battisti}, Andrew J. and {Brammer}, Gabriel and {Cai}, Zheng and {Champagne}, Jaclyn and {Drakos}, Nicole E. and {Egami}, Eiichi and {Fan}, Xiaohui and {Franco}, Maximilien and {Fudamoto}, Yoshinobu and {Fujimoto}, Seiji and {Gillman}, Steven and {Gozaliasl}, Ghassem and {Harish}, Santosh and {Jin}, Xiangyu and {Kakiichi}, Koki and {Kakkad}, Darshan and {Koekemoer}, Anton M. and {Lin}, Ruqiu and {Liu}, Daizhong and {Long}, Arianna S. and {Magdis}, Georgios E. and {Manning}, Sinclaire and {Martin}, Crystal L. and {McKinney}, Jed and {Meyer}, Romain and {Rodighiero}, Giulia and {Salazar}, Victoria and {Sanders}, David B. and {Shuntov}, Marko and {Talia}, Margherita and {Tanaka}, Takumi S. and {Wang}, Feige and {Wang}, Wuji and {Wilkins}, Stephen M. and {Yang}, Jinyi and {Yun}, Min S. and {The Champs} and {Cosmos-Web Collaborations}},
        title = "{ALMA and JWST Identification of Faint Dusty Star-forming Galaxies up to z {\ensuremath{\sim}} 8 and Their Connection with Other Galaxy Populations}",
      journal = {\apjl},
         year = 2026,
        month = feb,
       volume = {998},
       number = {2},
          eid = {L36},
        pages = {L36},
          doi = {10.3847/2041-8213/ae382a},
archivePrefix = {arXiv},
       eprint = {2512.16215},
 primaryClass = {astro-ph.GA},
       adsurl = {https://ui.adsabs.harvard.edu/abs/2026ApJ...998L..36Z}
}

@BOOK{1998asls.book.....B,
       author = {{Bohren}, Craig F. and {Huffman}, Donald R.},
        title = "{Absorption and Scattering of Light by Small Particles}",
         year = 1998,
       adsurl = {https://ui.adsabs.harvard.edu/abs/1998asls.book.....B}
}

@software{gomez_munoz_2026_20795378,
  author       = {Gómez-Muñoz, Marco A.},
  title        = {MergerCurve: A tool for transient light curve
                   interpolation
                  },
  month        = jun,
  year         = 2026,
  publisher    = {Zenodo},
  doi          = {10.5281/zenodo.20795378},
  url          = {https://doi.org/10.5281/zenodo.20795378},
}

@ARTICLE{2006A&A...460..339J,
       author = {{Jordi}, K. and {Grebel}, E.~K. and {Ammon}, K.},
        title = "{Empirical color transformations between SDSS photometry and other photometric systems}",
      journal = {\aap},
         year = 2006,
        month = dec,
       volume = {460},
       number = {1},
        pages = {339-347},
          doi = {10.1051/0004-6361:20066082},
archivePrefix = {arXiv},
       eprint = {astro-ph/0609121},
 primaryClass = {astro-ph},
       adsurl = {https://ui.adsabs.harvard.edu/abs/2006A&A...460..339J}
}

@ARTICLE{2016A&A...595A...1G,
       author = {{Gaia Collaboration} and {Prusti}, T. and {de Bruijne}, J.~H.~J. and {Brown}, A.~G.~A. and {Vallenari}, A. and {Babusiaux}, C. and {Bailer-Jones}, C.~A.~L. and {Bastian}, U. and {Biermann}, M. and {Evans}, D.~W. and {Eyer}, L. and {Jansen}, F. and {Jordi}, C. and {Klioner}, S.~A. and {Lammers}, U. and {Lindegren}, L. and {Luri}, X. and {Mignard}, F. and {Milligan}, D.~J. and {Panem}, C. and {Poinsignon}, V. and {Pourbaix}, D. and {Randich}, S. and {Sarri}, G. and {Sartoretti}, P. and {Siddiqui}, H.~I. and {Soubiran}, C. and {Valette}, V. and {van Leeuwen}, F. and {Walton}, N.~A. and {Aerts}, C. and {Arenou}, F. and {Cropper}, M. and {Drimmel}, R. and {H{\o}g}, E. and {Katz}, D. and {Lattanzi}, M.~G. and {O'Mullane}, W. and {Grebel}, E.~K. and {Holland}, A.~D. and {Huc}, C. and {Passot}, X. and {Bramante}, L. and {Cacciari}, C. and {Casta{\~n}eda}, J. and {Chaoul}, L. and {Cheek}, N. and {De Angeli}, F. and {Fabricius}, C. and {Guerra}, R. and {Hern{\'a}ndez}, J. and {Jean-Antoine-Piccolo}, A. and {Masana}, E. and {Messineo}, R. and {Mowlavi}, N. and {Nienartowicz}, K. and {Ord{\'o}{\~n}ez-Blanco}, D. and {Panuzzo}, P. and {Portell}, J. and {Richards}, P.~J. and {Riello}, M. and {Seabroke}, G.~M. and {Tanga}, P. and {Th{\'e}venin}, F. and {Torra}, J. and {Els}, S.~G. and {Gracia-Abril}, G. and {Comoretto}, G. and {Garcia-Reinaldos}, M. and {Lock}, T. and {Mercier}, E. and {Altmann}, M. and {Andrae}, R. and {Astraatmadja}, T.~L. and {Bellas-Velidis}, I. and {Benson}, K. and {Berthier}, J. and {Blomme}, R. and {Busso}, G. and {Carry}, B. and {Cellino}, A. and {Clementini}, G. and {Cowell}, S. and {Creevey}, O. and {Cuypers}, J. and {Davidson}, M. and {De Ridder}, J. and {de Torres}, A. and {Delchambre}, L. and {Dell'Oro}, A. and {Ducourant}, C. and {Fr{\'e}mat}, Y. and {Garc{\'\i}a-Torres}, M. and {Gosset}, E. and {Halbwachs}, J.-L. and {Hambly}, N.~C. and {Harrison}, D.~L. and {Hauser}, M. and {Hestroffer}, D. and {Hodgkin}, S.~T. and {Huckle}, H.~E. and {Hutton}, A. and {Jasniewicz}, G. and {Jordan}, S. and {Kontizas}, M. and {Korn}, A.~J. and {Lanzafame}, A.~C. and {Manteiga}, M. and {Moitinho}, A. and {Muinonen}, K. and {Osinde}, J. and {Pancino}, E. and {Pauwels}, T. and {Petit}, J.-M. and {Recio-Blanco}, A. and {Robin}, A.~C. and {Sarro}, L.~M. and {Siopis}, C. and {Smith}, M. and {Smith}, K.~W. and {Sozzetti}, A. and {Thuillot}, W. and {van Reeven}, W. and {Viala}, Y. and {Abbas}, U. and {Abreu Aramburu}, A. and {Accart}, S. and {Aguado}, J.~J. and {Allan}, P.~M. and {Allasia}, W. and {Altavilla}, G. and {{\'A}lvarez}, M.~A. and {Alves}, J. and {Anderson}, R.~I. and {Andrei}, A.~H. and {Anglada Varela}, E. and {Antiche}, E. and {Antoja}, T. and {Ant{\'o}n}, S. and {Arcay}, B. and {Atzei}, A. and {Ayache}, L. and {Bach}, N. and {Baker}, S.~G. and {Balaguer-N{\'u}{\~n}ez}, L. and {Barache}, C. and {Barata}, C. and {Barbier}, A. and {Barblan}, F. and {Baroni}, M. and {Barrado y Navascu{\'e}s}, D. and {Barros}, M. and {Barstow}, M.~A. and {Becciani}, U. and {Bellazzini}, M. and {Bellei}, G. and {Bello Garc{\'\i}a}, A. and {Belokurov}, V. and {Bendjoya}, P. and {Berihuete}, A. and {Bianchi}, L. and {Bienaym{\'e}}, O. and {Billebaud}, F. and {Blagorodnova}, N. and {Blanco-Cuaresma}, S. and {Boch}, T. and {Bombrun}, A. and {Borrachero}, R. and {Bouquillon}, S. and {Bourda}, G. and {Bouy}, H. and {Bragaglia}, A. and {Breddels}, M.~A. and {Brouillet}, N. and {Br{\"u}semeister}, T. and {Bucciarelli}, B. and {Budnik}, F. and {Burgess}, P. and {Burgon}, R. and {Burlacu}, A. and {Busonero}, D. and {Buzzi}, R. and {Caffau}, E. and {Cambras}, J. and {Campbell}, H. and {Cancelliere}, R. and {Cantat-Gaudin}, T. and {Carlucci}, T. and {Carrasco}, J.~M. and {Castellani}, M. and {Charlot}, P. and {Charnas}, J. and {Charvet}, P. and {Chassat}, F. and {Chiavassa}, A. and {Clotet}, M. and {Cocozza}, G. and {Collins}, R.~S. and {Collins}, P. and {Costigan}, G.},
        title = "{The Gaia mission}",
      journal = {\aap},
         year = 2016,
        month = nov,
       volume = {595},
          eid = {A1},
        pages = {A1},
          doi = {10.1051/0004-6361/201629272},
archivePrefix = {arXiv},
       eprint = {1609.04153},
 primaryClass = {astro-ph.IM},
       adsurl = {https://ui.adsabs.harvard.edu/abs/2016A&A...595A...1G}
}

@ARTICLE{2023A&A...674A...1G,
       author = {{Gaia Collaboration} and {Vallenari}, A. and {Brown}, A.~G.~A. and {Prusti}, T. and {de Bruijne}, J.~H.~J. and {Arenou}, F. and {Babusiaux}, C. and {Biermann}, M. and {Creevey}, O.~L. and {Ducourant}, C. and {Evans}, D.~W. and {Eyer}, L. and {Guerra}, R. and {Hutton}, A. and {Jordi}, C. and {Klioner}, S.~A. and {Lammers}, U.~L. and {Lindegren}, L. and {Luri}, X. and {Mignard}, F. and {Panem}, C. and {Pourbaix}, D. and {Randich}, S. and {Sartoretti}, P. and {Soubiran}, C. and {Tanga}, P. and {Walton}, N.~A. and {Bailer-Jones}, C.~A.~L. and {Bastian}, U. and {Drimmel}, R. and {Jansen}, F. and {Katz}, D. and {Lattanzi}, M.~G. and {van Leeuwen}, F. and {Bakker}, J. and {Cacciari}, C. and {Casta{\~n}eda}, J. and {De Angeli}, F. and {Fabricius}, C. and {Fouesneau}, M. and {Fr{\'e}mat}, Y. and {Galluccio}, L. and {Guerrier}, A. and {Heiter}, U. and {Masana}, E. and {Messineo}, R. and {Mowlavi}, N. and {Nicolas}, C. and {Nienartowicz}, K. and {Pailler}, F. and {Panuzzo}, P. and {Riclet}, F. and {Roux}, W. and {Seabroke}, G.~M. and {Sordo}, R. and {Th{\'e}venin}, F. and {Gracia-Abril}, G. and {Portell}, J. and {Teyssier}, D. and {Altmann}, M. and {Andrae}, R. and {Audard}, M. and {Bellas-Velidis}, I. and {Benson}, K. and {Berthier}, J. and {Blomme}, R. and {Burgess}, P.~W. and {Busonero}, D. and {Busso}, G. and {C{\'a}novas}, H. and {Carry}, B. and {Cellino}, A. and {Cheek}, N. and {Clementini}, G. and {Damerdji}, Y. and {Davidson}, M. and {de Teodoro}, P. and {Nu{\~n}ez Campos}, M. and {Delchambre}, L. and {Dell'Oro}, A. and {Esquej}, P. and {Fern{\'a}ndez-Hern{\'a}ndez}, J. and {Fraile}, E. and {Garabato}, D. and {Garc{\'\i}a-Lario}, P. and {Gosset}, E. and {Haigron}, R. and {Halbwachs}, J.-L. and {Hambly}, N.~C. and {Harrison}, D.~L. and {Hern{\'a}ndez}, J. and {Hestroffer}, D. and {Hodgkin}, S.~T. and {Holl}, B. and {Jan{\ss}en}, K. and {Jevardat de Fombelle}, G. and {Jordan}, S. and {Krone-Martins}, A. and {Lanzafame}, A.~C. and {L{\"o}ffler}, W. and {Marchal}, O. and {Marrese}, P.~M. and {Moitinho}, A. and {Muinonen}, K. and {Osborne}, P. and {Pancino}, E. and {Pauwels}, T. and {Recio-Blanco}, A. and {Reyl{\'e}}, C. and {Riello}, M. and {Rimoldini}, L. and {Roegiers}, T. and {Rybizki}, J. and {Sarro}, L.~M. and {Siopis}, C. and {Smith}, M. and {Sozzetti}, A. and {Utrilla}, E. and {van Leeuwen}, M. and {Abbas}, U. and {{\'A}brah{\'a}m}, P. and {Abreu Aramburu}, A. and {Aerts}, C. and {Aguado}, J.~J. and {Ajaj}, M. and {Aldea-Montero}, F. and {Altavilla}, G. and {{\'A}lvarez}, M.~A. and {Alves}, J. and {Anders}, F. and {Anderson}, R.~I. and {Anglada Varela}, E. and {Antoja}, T. and {Baines}, D. and {Baker}, S.~G. and {Balaguer-N{\'u}{\~n}ez}, L. and {Balbinot}, E. and {Balog}, Z. and {Barache}, C. and {Barbato}, D. and {Barros}, M. and {Barstow}, M.~A. and {Bartolom{\'e}}, S. and {Bassilana}, J.-L. and {Bauchet}, N. and {Becciani}, U. and {Bellazzini}, M. and {Berihuete}, A. and {Bernet}, M. and {Bertone}, S. and {Bianchi}, L. and {Binnenfeld}, A. and {Blanco-Cuaresma}, S. and {Blazere}, A. and {Boch}, T. and {Bombrun}, A. and {Bossini}, D. and {Bouquillon}, S. and {Bragaglia}, A. and {Bramante}, L. and {Breedt}, E. and {Bressan}, A. and {Brouillet}, N. and {Brugaletta}, E. and {Bucciarelli}, B. and {Burlacu}, A. and {Butkevich}, A.~G. and {Buzzi}, R. and {Caffau}, E. and {Cancelliere}, R. and {Cantat-Gaudin}, T. and {Carballo}, R. and {Carlucci}, T. and {Carnerero}, M.~I. and {Carrasco}, J.~M. and {Casamiquela}, L. and {Castellani}, M. and {Castro-Ginard}, A. and {Chaoul}, L. and {Charlot}, P. and {Chemin}, L. and {Chiaramida}, V. and {Chiavassa}, A. and {Chornay}, N. and {Comoretto}, G. and {Contursi}, G. and {Cooper}, W.~J. and {Cornez}, T. and {Cowell}, S. and {Crifo}, F. and {Cropper}, M. and {Crosta}, M. and {Crowley}, C. and {Dafonte}, C. and {Dapergolas}, A. and {David}, M. and {David}, P. and {de Laverny}, P. and {De Luise}, F. and {De March}, R.},
        title = "{Gaia Data Release 3. Summary of the content and survey properties}",
      journal = {\aap},
         year = 2023,
        month = jun,
       volume = {674},
          eid = {A1},
        pages = {A1},
          doi = {10.1051/0004-6361/202243940},
archivePrefix = {arXiv},
       eprint = {2208.00211},
 primaryClass = {astro-ph.GA},
       adsurl = {https://ui.adsabs.harvard.edu/abs/2023A&A...674A...1G}
}

@ARTICLE{2019ApJ...873..111I,
       author = {{Ivezi{\'c}}, {\v{Z}}eljko and {Kahn}, Steven M. and {Tyson}, J. Anthony and {Abel}, Bob and {Acosta}, Emily and {Allsman}, Robyn and {Alonso}, David and {AlSayyad}, Yusra and {Anderson}, Scott F. and {Andrew}, John and {Angel}, James Roger P. and {Angeli}, George Z. and {Ansari}, Reza and {Antilogus}, Pierre and {Araujo}, Constanza and {Armstrong}, Robert and {Arndt}, Kirk T. and {Astier}, Pierre and {Aubourg}, {\'E}ric and {Auza}, Nicole and {Axelrod}, Tim S. and {Bard}, Deborah J. and {Barr}, Jeff D. and {Barrau}, Aurelian and {Bartlett}, James G. and {Bauer}, Amanda E. and {Bauman}, Brian J. and {Baumont}, Sylvain and {Bechtol}, Ellen and {Bechtol}, Keith and {Becker}, Andrew C. and {Becla}, Jacek and {Beldica}, Cristina and {Bellavia}, Steve and {Bianco}, Federica B. and {Biswas}, Rahul and {Blanc}, Guillaume and {Blazek}, Jonathan and {Blandford}, Roger D. and {Bloom}, Josh S. and {Bogart}, Joanne and {Bond}, Tim W. and {Booth}, Michael T. and {Borgland}, Anders W. and {Borne}, Kirk and {Bosch}, James F. and {Boutigny}, Dominique and {Brackett}, Craig A. and {Bradshaw}, Andrew and {Brandt}, William Nielsen and {Brown}, Michael E. and {Bullock}, James S. and {Burchat}, Patricia and {Burke}, David L. and {Cagnoli}, Gianpietro and {Calabrese}, Daniel and {Callahan}, Shawn and {Callen}, Alice L. and {Carlin}, Jeffrey L. and {Carlson}, Erin L. and {Chandrasekharan}, Srinivasan and {Charles-Emerson}, Glenaver and {Chesley}, Steve and {Cheu}, Elliott C. and {Chiang}, Hsin-Fang and {Chiang}, James and {Chirino}, Carol and {Chow}, Derek and {Ciardi}, David R. and {Claver}, Charles F. and {Cohen-Tanugi}, Johann and {Cockrum}, Joseph J. and {Coles}, Rebecca and {Connolly}, Andrew J. and {Cook}, Kem H. and {Cooray}, Asantha and {Covey}, Kevin R. and {Cribbs}, Chris and {Cui}, Wei and {Cutri}, Roc and {Daly}, Philip N. and {Daniel}, Scott F. and {Daruich}, Felipe and {Daubard}, Guillaume and {Daues}, Greg and {Dawson}, William and {Delgado}, Francisco and {Dellapenna}, Alfred and {de Peyster}, Robert and {de Val-Borro}, Miguel and {Digel}, Seth W. and {Doherty}, Peter and {Dubois}, Richard and {Dubois-Felsmann}, Gregory P. and {Durech}, Josef and {Economou}, Frossie and {Eifler}, Tim and {Eracleous}, Michael and {Emmons}, Benjamin L. and {Fausti Neto}, Angelo and {Ferguson}, Henry and {Figueroa}, Enrique and {Fisher-Levine}, Merlin and {Focke}, Warren and {Foss}, Michael D. and {Frank}, James and {Freemon}, Michael D. and {Gangler}, Emmanuel and {Gawiser}, Eric and {Geary}, John C. and {Gee}, Perry and {Geha}, Marla and {Gessner}, Charles J.~B. and {Gibson}, Robert R. and {Gilmore}, D. Kirk and {Glanzman}, Thomas and {Glick}, William and {Goldina}, Tatiana and {Goldstein}, Daniel A. and {Goodenow}, Iain and {Graham}, Melissa L. and {Gressler}, William J. and {Gris}, Philippe and {Guy}, Leanne P. and {Guyonnet}, Augustin and {Haller}, Gunther and {Harris}, Ron and {Hascall}, Patrick A. and {Haupt}, Justine and {Hernandez}, Fabio and {Herrmann}, Sven and {Hileman}, Edward and {Hoblitt}, Joshua and {Hodgson}, John A. and {Hogan}, Craig and {Howard}, James D. and {Huang}, Dajun and {Huffer}, Michael E. and {Ingraham}, Patrick and {Innes}, Walter R. and {Jacoby}, Suzanne H. and {Jain}, Bhuvnesh and {Jammes}, Fabrice and {Jee}, M. James and {Jenness}, Tim and {Jernigan}, Garrett and {Jevremovi{\'c}}, Darko and {Johns}, Kenneth and {Johnson}, Anthony S. and {Johnson}, Margaret W.~G. and {Jones}, R. Lynne and {Juramy-Gilles}, Claire and {Juri{\'c}}, Mario and {Kalirai}, Jason S. and {Kallivayalil}, Nitya J. and {Kalmbach}, Bryce and {Kantor}, Jeffrey P. and {Karst}, Pierre and {Kasliwal}, Mansi M. and {Kelly}, Heather and {Kessler}, Richard and {Kinnison}, Veronica and {Kirkby}, David and {Knox}, Lloyd and {Kotov}, Ivan V. and {Krabbendam}, Victor L. and {Krughoff}, K. Simon and {Kub{\'a}nek}, Petr and {Kuczewski}, John and {Kulkarni}, Shri and {Ku}, John and {Kurita}, Nadine R. and {Lage}, Craig S. and {Lambert}, Ron and {Lange}, Travis and {Langton}, J. Brian and {Le Guillou}, Laurent and {Levine}, Deborah and {Liang}, Ming and {Lim}, Kian-Tat and {Lintott}, Chris J. and {Long}, Kevin E. and {Lopez}, Margaux and {Lotz}, Paul J. and {Lupton}, Robert H. and {Lust}, Nate B. and {MacArthur}, Lauren A. and {Mahabal}, Ashish and {Mandelbaum}, Rachel and {Markiewicz}, Thomas W. and {Marsh}, Darren S. and {Marshall}, Philip J. and {Marshall}, Stuart and {May}, Morgan and {McKercher}, Robert and {McQueen}, Michelle and {Meyers}, Joshua and {Migliore}, Myriam and {Miller}, Michelle and {Mills}, David J.},
        title = "{LSST: From Science Drivers to Reference Design and Anticipated Data Products}",
      journal = {\apj},
         year = 2019,
        month = mar,
       volume = {873},
       number = {2},
          eid = {111},
        pages = {111},
          doi = {10.3847/1538-4357/ab042c},
archivePrefix = {arXiv},
       eprint = {0805.2366},
 primaryClass = {astro-ph},
       adsurl = {https://ui.adsabs.harvard.edu/abs/2019ApJ...873..111I}
}

@ARTICLE{2017ApJS..230...15M,
       author = {{Moe}, Maxwell and {Di Stefano}, Rosanne},
        title = "{Mind Your Ps and Qs: The Interrelation between Period (P) and Mass-ratio (Q) Distributions of Binary Stars}",
      journal = {\apjs},
         year = 2017,
        month = jun,
       volume = {230},
       number = {2},
          eid = {15},
        pages = {15},
          doi = {10.3847/1538-4365/aa6fb6},
archivePrefix = {arXiv},
       eprint = {1606.05347},
 primaryClass = {astro-ph.SR},
       adsurl = {https://ui.adsabs.harvard.edu/abs/2017ApJS..230...15M}
}

@ARTICLE{1989ApJ...341..867C,
       author = {{Chevalier}, Roger A. and {Soker}, Noam},
        title = "{Asymmetric Envelope Expansion of Supernova 1987A}",
      journal = {\apj},
         year = 1989,
        month = jun,
       volume = {341},
        pages = {867},
          doi = {10.1086/167545},
       adsurl = {https://ui.adsabs.harvard.edu/abs/1989ApJ...341..867C}
}

@ARTICLE{2026arXiv260622619W,
       author = {{Wavasseur}, M. and {Blagorodnova}, N. and {G{\'o}mez-Mu{\~n}oz}, M.~A. and {Pols}, O.~R.},
        title = "{Comparative Study of Two Luminous Red Novae I. Progenitor Modeling and Dust Formation}",
      journal = {arXiv e-prints},
         year = 2026,
        month = jun,
          eid = {arXiv:2606.22619},
        pages = {arXiv:2606.22619},
          doi = {10.48550/arXiv.2606.22619},
archivePrefix = {arXiv},
       eprint = {2606.22619},
 primaryClass = {astro-ph.SR},
       adsurl = {https://ui.adsabs.harvard.edu/abs/2026arXiv260622619W}
}

@ARTICLE{2013A&A...558A..82K,
       author = {{Kami{\'n}ski}, T. and {Tylenda}, R.},
        title = "{Optical spectropolarimetry of V4332 Sagittarii}",
      journal = {\aap},
         year = 2013,
        month = oct,
       volume = {558},
          eid = {A82},
        pages = {A82},
          doi = {10.1051/0004-6361/201321852},
archivePrefix = {arXiv},
       eprint = {1309.0635},
 primaryClass = {astro-ph.SR},
       adsurl = {https://ui.adsabs.harvard.edu/abs/2013A&A...558A..82K}
}

@ARTICLE{2023Natur.617...55D,
       author = {{De}, Kishalay and {MacLeod}, Morgan and {Karambelkar}, Viraj and {Jencson}, Jacob E. and {Chakrabarty}, Deepto and {Conroy}, Charlie and {Dekany}, Richard and {Eilers}, Anna-Christina and {Graham}, Matthew J. and {Hillenbrand}, Lynne A. and {Kara}, Erin and {Kasliwal}, Mansi M. and {Kulkarni}, S.~R. and {Lau}, Ryan M. and {Loeb}, Abraham and {Masci}, Frank and {Medford}, Michael S. and {Meisner}, Aaron M. and {Patel}, Nimesh and {Quiroga-Nu{\~n}ez}, Luis Henry and {Riddle}, Reed L. and {Rusholme}, Ben and {Simcoe}, Robert and {Sjouwerman}, Lor{\'a}nt O. and {Teague}, Richard and {Vanderburg}, Andrew},
        title = "{An infrared transient from a star engulfing a planet}",
      journal = {\nat},
         year = 2023,
        month = may,
       volume = {617},
       number = {7959},
        pages = {55-60},
          doi = {10.1038/s41586-023-05842-x},
       adsurl = {https://ui.adsabs.harvard.edu/abs/2023Natur.617...55D}
}

@ARTICLE{2022ApJ...938....5M,
       author = {{Matsumoto}, Tatsuya and {Metzger}, Brian D.},
        title = "{Light-curve Model for Luminous Red Novae and Inferences about the Ejecta of Stellar Mergers}",
      journal = {\apj},
         year = 2022,
        month = oct,
       volume = {938},
       number = {1},
          eid = {5},
        pages = {5},
          doi = {10.3847/1538-4357/ac6269},
archivePrefix = {arXiv},
       eprint = {2202.10478},
 primaryClass = {astro-ph.SR},
       adsurl = {https://ui.adsabs.harvard.edu/abs/2022ApJ...938....5M}
}

@ARTICLE{2025A&A...699A.316S,
       author = {{Steinmetz}, T. and {Kami{\'n}ski}, T. and {Melis}, C. and {Blagorodnova}, N. and {Gromadzki}, M. and {Menten}, K. and {Su}, K.},
        title = "{OGLE-2002-BLG-360: A dusty anomaly among red nova remnants}",
      journal = {\aap},
         year = 2025,
        month = jul,
       volume = {699},
          eid = {A316},
        pages = {A316},
          doi = {10.1051/0004-6361/202554261},
archivePrefix = {arXiv},
       eprint = {2502.18365},
 primaryClass = {astro-ph.SR},
       adsurl = {https://ui.adsabs.harvard.edu/abs/2025A&A...699A.316S}
}

@ARTICLE{2014ApJ...788...22P,
       author = {{Pejcha}, Ond{\v{r}}ej},
        title = "{Burying a Binary: Dynamical Mass Loss and a Continuous Optically thick Outflow Explain the Candidate Stellar Merger V1309 Scorpii}",
      journal = {\apj},
         year = 2014,
        month = jun,
       volume = {788},
       number = {1},
          eid = {22},
        pages = {22},
          doi = {10.1088/0004-637X/788/1/22},
archivePrefix = {arXiv},
       eprint = {1307.4088},
 primaryClass = {astro-ph.SR},
       adsurl = {https://ui.adsabs.harvard.edu/abs/2014ApJ...788...22P}
}

@misc{aplpy2012,
  author        = {{Robitaille}, T. and {Bressert}, E.},
  title         = "{APLpy: Astronomical Plotting Library in Python}",
  howpublished  = {Astrophysics Source Code Library},
  year          = 2012,
  month         = aug,
  archivePrefix = "ascl",
  eprint        = {1208.017},
  adsurl        = {http://adsabs.harvard.edu/abs/2012ascl.soft08017R}
}

@misc{aplpy2019,
  author       = {Robitaille, Thomas},
  title        = {{APLpy v2.0: The Astronomical Plotting Library in Python}},
  month        = feb,
  year         = 2019,
  doi          = {10.5281/zenodo.2567476},
  url          = {https://doi.org/10.5281/zenodo.2567476}
}

@ARTICLE{2016A&A...592A.134T,
       author = {{Tylenda}, R. and {Kami{\'n}ski}, T.},
        title = "{Evolution of the stellar-merger red nova V1309 Scorpii: Spectral energy distribution analysis}",
      journal = {\aap},
         year = 2016,
        month = aug,
       volume = {592},
          eid = {A134},
        pages = {A134},
          doi = {10.1051/0004-6361/201527700},
archivePrefix = {arXiv},
       eprint = {1606.09426},
 primaryClass = {astro-ph.SR},
       adsurl = {https://ui.adsabs.harvard.edu/abs/2016A&A...592A.134T}
}

@article{pypeit:joss_pub,
    doi = {10.21105/joss.02308},
    url = {https://doi.org/10.21105/joss.02308},
    year = {2020},
    publisher = {The Open Journal},
    volume = {5},
    number = {56},
    pages = {2308},
    author = {J. Xavier Prochaska and Joseph F. Hennawi and Kyle B. Westfall and Ryan J. Cooke and Feige Wang and Tiffany Hsyu and Frederick B. Davies and Emanuele Paolo Farina and Debora Pelliccia},
    title = {PypeIt: The Python Spectroscopic Data Reduction Pipeline},
    journal = {Journal of Open Source Software}
}

@MISC{pypeit:zenodo,
       author = {{Prochaska}, J. Xavier and {Hennawi}, Joseph and {Cooke}, Ryan and
         {Westfall}, Kyle and {Wang}, Feige and {EmAstro} and {Tiffanyhsyu} and
         {Wasserman}, Asher and {Villaume}, Alexa and {Marijana777} and
         {Schindler}, JT and {Young}, David and {Simha}, Sunil and
         {Wilde}, Matt and {Tejos}, Nicolas and {Isbell}, Jacob and
         {Fl{\"o}rs}, Andreas and {Sandford}, Nathan and {Vasovi{\'c}}, Zlatan and
         {Betts}, Edward and {Holden}, Brad},
        title = "{pypeit/PypeIt: Release 1.0.0}",
         year = 2020,
        month = apr,
          eid = {10.5281/zenodo.3743493},
          doi = {10.5281/zenodo.3743493},
      version = {v1.0.0},
    publisher = {Zenodo},
       adsurl = {https://ui.adsabs.harvard.edu/abs/2020zndo...3743493P}
}

@ARTICLE{2024ApJ...961..211M,
       author = {{Masterson}, Megan and {De}, Kishalay and {Panagiotou}, Christos and {Kara}, Erin and {Arcavi}, Iair and {Eilers}, Anna-Christina and {Frostig}, Danielle and {Gezari}, Suvi and {Grotova}, Iuliia and {Liu}, Zhu and et al.},
        title = "{A New Population of Mid-infrared-selected Tidal Disruption Events: Implications for Tidal Disruption Event Rates and Host Galaxy Properties}",
      journal = {\apj},
         year = 2024,
        month = feb,
       volume = {961},
       number = {2},
          eid = {211},
        pages = {211},
          doi = {10.3847/1538-4357/ad18bb},
archivePrefix = {arXiv},
       eprint = {2401.01403},
 primaryClass = {astro-ph.HE},
       adsurl = {https://ui.adsabs.harvard.edu/abs/2024ApJ...961..211M}
}

@ARTICLE{2011ApJ...731...53M,
       author = {{Mainzer}, A. and {Bauer}, J. and {Grav}, T. and {Masiero}, J. and {Cutri}, R.~M. and {Dailey}, J. and {Eisenhardt}, P. and {McMillan}, R.~S. and {Wright}, E. and {Walker}, R. and et al.},
        title = "{Preliminary Results from NEOWISE: An Enhancement to the Wide-field Infrared Survey Explorer for Solar System Science}",
      journal = {\apj},
         year = 2011,
        month = apr,
       volume = {731},
       number = {1},
          eid = {53},
        pages = {53},
          doi = {10.1088/0004-637X/731/1/53},
archivePrefix = {arXiv},
       eprint = {1102.1996},
 primaryClass = {astro-ph.EP},
       adsurl = {https://ui.adsabs.harvard.edu/abs/2011ApJ...731...53M}
}

@ARTICLE{2014ApJ...792...30M,
       author = {{Mainzer}, A. and {Bauer}, J. and {Cutri}, R.~M. and {Grav}, T. and {Masiero}, J. and {Beck}, R. and {Clarkson}, P. and {Conrow}, T. and {Dailey}, J. and {Eisenhardt}, P. and et al.},
        title = "{Initial Performance of the NEOWISE Reactivation Mission}",
      journal = {\apj},
         year = 2014,
        month = sep,
       volume = {792},
       number = {1},
          eid = {30},
        pages = {30},
          doi = {10.1088/0004-637X/792/1/30},
archivePrefix = {arXiv},
       eprint = {1406.6025},
 primaryClass = {astro-ph.EP},
       adsurl = {https://ui.adsabs.harvard.edu/abs/2014ApJ...792...30M}
}

@ARTICLE{2016ApJ...830...27Z,
       author = {{Zackay}, Barak and {Ofek}, Eran O. and {Gal-Yam}, Avishay},
        title = "{Proper Image Subtraction{\textemdash}Optimal Transient Detection, Photometry, and Hypothesis Testing}",
      journal = {\apj},
         year = 2016,
        month = oct,
       volume = {830},
       number = {1},
          eid = {27},
        pages = {27},
          doi = {10.3847/0004-637X/830/1/27},
archivePrefix = {arXiv},
       eprint = {1601.02655},
 primaryClass = {astro-ph.IM},
       adsurl = {https://ui.adsabs.harvard.edu/abs/2016ApJ...830...27Z}
}

@INPROCEEDINGS{2012SPIE.8446E..0JM,
       author = {{McLean}, Ian S. and {Steidel}, Charles C. and {Epps}, Harland W. and {Konidaris}, Nicholas and {Matthews}, Keith Y. and {Adkins}, Sean and {Aliado}, Theodore and {Brims}, George and {Canfield}, John M. and {Cromer}, John L. and et al.},
        title = "{MOSFIRE, the multi-object spectrometer for infra-red exploration at the Keck Observatory}",
    booktitle = {Ground-based and Airborne Instrumentation for Astronomy IV},
         year = 2012,
       editor = {{McLean}, Ian S. and {Ramsay}, Suzanne K. and {Takami}, Hideki},
       series = {Society of Photo-Optical Instrumentation Engineers (SPIE) Conference Series},
       volume = {8446},
        month = sep,
          eid = {84460J},
        pages = {84460J},
          doi = {10.1117/12.924794},
       adsurl = {https://ui.adsabs.harvard.edu/abs/2012SPIE.8446E..0JM}
}

@ARTICLE{2023ApJ...955L...8H,
       author = {{Hiramatsu}, Daichi and {Tsuna}, Daichi and {Berger}, Edo and {Itagaki}, Koichi and {Goldberg}, Jared A. and {Gomez}, Sebastian and {Kishalay}, De and {Hosseinzadeh}, Griffin and {Bostroem}, K. Azalee and {Brown}, Peter J. and {Arcavi}, Iair and {Bieryla}, Allyson and {Blanchard}, Peter K. and {Esquerdo}, Gilbert A. and {Farah}, Joseph and {Howell}, D. Andrew and {Matsumoto}, Tatsuya and {McCully}, Curtis and {Newsome}, Megan and {Gonzalez}, Estefania Padilla and {Pellegrino}, Craig and {Rhee}, Jaehyon and {Terreran}, Giacomo and {Vink{\'o}}, J{\'o}zsef and {Wheeler}, J. Craig},
        title = "{From Discovery to the First Month of the Type II Supernova 2023ixf: High and Variable Mass Loss in the Final Year before Explosion}",
      journal = {\apjl},
         year = 2023,
        month = sep,
       volume = {955},
       number = {1},
          eid = {L8},
        pages = {L8},
          doi = {10.3847/2041-8213/acf299},
archivePrefix = {arXiv},
       eprint = {2307.03165},
 primaryClass = {astro-ph.HE},
       adsurl = {https://ui.adsabs.harvard.edu/abs/2023ApJ...955L...8H}
}

\begin{appendix}

    \section{Observing logs and MIR light curve data}
    \label{ap:photometry_table}

\begin{figure*}
    \centering
    \includegraphics[width=0.8\linewidth]{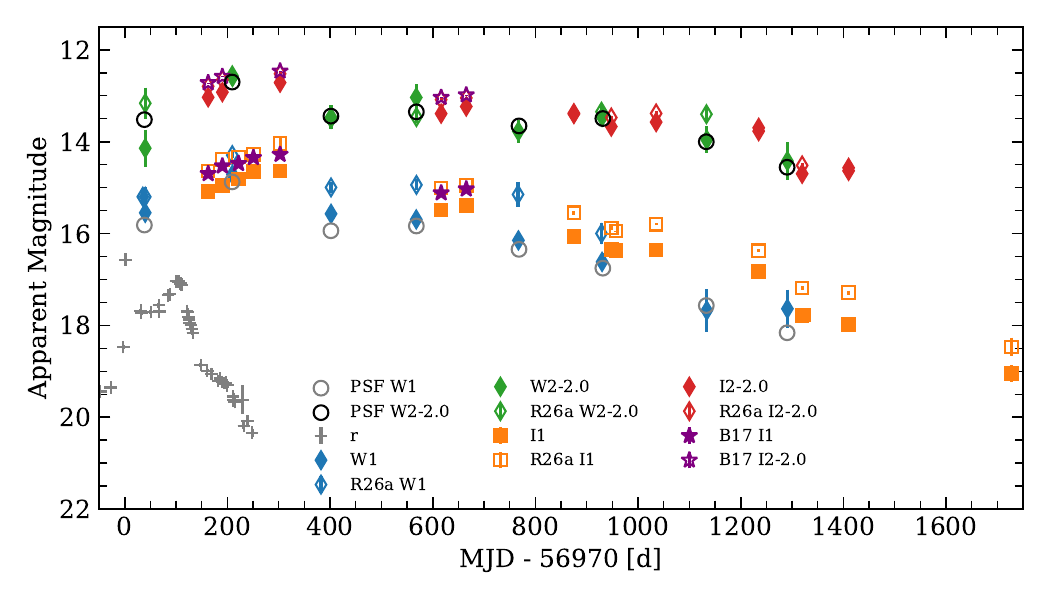}
    \caption{MIR Spitzer and NEOWISE light curves comparison of M101-OT as obtained in this work, those from \citet[][]{2026A&A...706A.154R}, R26a in the figure, and \citet{2017ApJ...834..107B}, B17 in the figure. The $r$-band light curve is set as reference. 
    \label{fig:B1}}
\end{figure*}

    Table~\ref{tab:log_obs} shows the observing logs of the NIRC2 and MOSFIRE photometric and spectral data.
    As mentioned in Sect.~\ref{sec:obs_data}, the M101-OT MIR photometry obtained from the NEOWISE mission and from the Spitzer space telescope were extracted using aperture photometry and corrected by the host galaxy contamination.
    The resulted MIR photometry, shown in Fig.~\ref{fig:B1}, is listed in Table~\ref{tab:photometry_table} along with measurements of the NIR from NIRC2 and MOSFIRE instruments. In this table, the Gaia DR3 multi-epoch photometry is also included. The optical photometry and the previously NIR photometry obtained from different instruments can be found in \citet{2017ApJ...834..107B}.

\begin{table*}
\caption{NIR and MIR photometry of the LRN M101-OT. \label{tab:photometry_table}}
\centering
 \begin{tabular}{lccccccccc}
\hline \hline
Instrument & Filter & MJD & $\sigma_\mathrm{min}^{MJD}$ & $\sigma_\mathrm{max}^{MJD}$ & Mag & $\sigma_\mathrm{Mag}$ & Flux & $\sigma_\mathrm{Flux}$ & Limit \\
 & & (d) & (d) & (d) & (mag) & (mag) & ($\mu$Jy) & ($\mu$Jy) & \\
\hline
  MOSFIRE & K & 58126.00 & .\,.\,. & .\,.\,. & 20.17 & 0.20 & 5.59 & 1.05 & 0 \\
  NIRC2 & K & 58303.49 & .\,.\,. & .\,.\,. & 21.38 & 0.12 & 1.83 & 0.20 & 0  \\
  NEOWISE & W1 & 57900.40 & 6.61 & 1.38 & 16.62 & 0.19 & 68.43 & 11.78 & 0  \\
  NEOWISE & W2 & 57180.40 & 0.70 & 0.81 & 14.59 & 0.11 & 245.73 & 26.00 & 0  \\
  Spitzer & I1 & 58005.29 & .\,.\,. & .\,.\,. & 16.36 & 0.01 & 78.56 & 0.75 & 0  \\
  Spitzer & I2 & 58005.29 & .\,.\,. & .\,.\,. & 15.57 & 0.01 & 104.89 & 0.83 & 0  \\
  GaiaDR3 & G & 56870.01 & .\,.\,. & .\,.\,. & 19.50 & 0.01 & 51.07 & 0.28 & 0  \\
  GaiaDR3 & Grp & 56872.26 & .\,.\,. & .\,.\,. & 19.31 & 0.01 & 48.34 & 0.27 & 0  \\
  GaiaDR3 & Gbp & 56942.26 & .\,.\,. & .\,.\,. & 19.55 & 0.01 & 53.66 & 0.30 & 0  \\
.\,.\,. & .\,.\,. & .\,.\,. & .\,.\,. & .\,.\,. & .\,.\,. & .\,.\,. & .\,.\,. \\
\hline
\end{tabular}
\tablefoot{
    Filters \textit{I1} and \textit{I2} refer to the $[3.6]$ and $[4.5]$ filters in Spitzer. Errors in MJD were only estimated for NEOWISE images and represent the extremes epochs of the co-added frames. \\ \\
    (This table is available in its entirety in machine-readable form.)
}
\end{table*}

\begin{table*}
    \caption{Log of imaging and spectroscopic observations. \label{tab:log_obs}}
    \centering
    \begin{tabular}{ccccccc}
    \hline \hline
       MJD  & Phase & Telescope/Instrument & Type\tablefootmark{a} & Filter & Exposure & Scale\tablefootmark{b} \\
       (d) & (d) &  &  &  & (s) & ({\arcsec}\,pix$^{-1}$)\\
    \hline
        57112 & 142 & KeckII/MOSFIRE & Spectrum & J & 2$\times$120 & 0.24 \\
        57112 & 142 & KeckII/MOSFIRE & Spectrum & H & 2$\times$120 & 0.24 \\
        57112 & 142 & KeckII/MOSFIRE & Spectrum & K & 4$\times$180 & 0.24 \\
        57167 & 197 & KeckII/MOSFIRE & Spectrum & Y & 4$\times$180 & 0.24 \\
        57167 & 197 & KeckII/MOSFIRE & Spectrum & J & 6$\times$120 & 0.24 \\
        57167 & 197 & KeckII/MOSFIRE & Spectrum & H & 6$\times$120 & 0.24 \\
        57167 & 197 & KeckII/MOSFIRE & Spectrum & K & 4$\times$180 & 0.24 \\
        57494 & 524 & KeckII/MOSFIRE & Spectrum & K & 6$\times$180 & 0.24 \\
        57538 & 568 & KeckII/MOSFIRE & Image & Ks & 9$\times$32 & 0.17\\
        57616 & 646 & KeckII/NIRC2 & Image & Ks & 3$\times$60 & 0.04 \\
        57848 & 878 & KeckII/NIRC2 & Image & H  & 2$\times$360 & 0.04 \\
        57848 & 878 & KeckII/NIRC2 & Image & Ks & 6$\times$300 & 0.04 \\
        58126 & 1156 & KeckII/MOSFIRE & Image & Ks & 6$\times$32 & 0.17\\
        58303 & 1333 & KeckII/NIRC2 & Image & Ks & 6$\times$300 & 0.04 \\
        58504 & 1534 & KeckII/NIRC2 & Image & Ks & 6$\times$300 & 0.04 \\
    \hline
    \end{tabular}
    \tablefoot{
    \tablefoottext{a}{All the MOSFIRE spectra were taken using a slit width of 0.7{\arcsec} with an instrumental resolving power R$\sim$3500.}
    \tablefoottext{b}{For MOSFIRE spectra the scale refers to the nominal scale in the dispersion direction.}
}
\end{table*}

\section{MIR photometry light curve comparison}
\label{ap:mir_photometry_comparison}

In Figure~\ref{fig:B1}, we compare our MIR Spitzer and NEOWISE photometry for M101-OT with the values published by \citet{2026A&A...706A.154R} and early Spitzer observations from \citet{2017ApJ...834..107B}. While there is close agreement (within $0.1$--$0.2$\,mag) between all studies in the redder MIR bands ($[4.5]$ and $W2$), our measurements in the bluest bands ($[3.6]$ and $W1$) are systematically fainter by $>0.5$\,mag compared to those studies.

In \citet{2026A&A...706A.154R}, they performed PSF-fitting photometry on image-subtracted frames using archival pre-outburst templates (March 2004 for Spitzer and June 2010 for NEOWISE; priv. comm.), removing the underlying host galaxy contribution. Similarly, \citet{2017ApJ...834..107B} employed template-subtracted PSF photometry for early Spitzer epochs. 

In our analysis, we implemented aperture photometry on seasonal co-added images combined with baseline flux subtraction for both NEOWISE and Spitzer. To test for potential host galaxy contamination in our aperture measurements (e.g. for local background over-subtraction), we also independently obtained PSF difference-imaging measurements \citep[based on the ZOGY algorithm][priv. comm.]{2016ApJ...830...27Z} that follows \citet{2023Natur.617...55D} \citep[see also][]{2024ApJ...961..211M}, as shown as open circles in Fig.~\ref{fig:B1}. These results show that the PSF difference-imaging photometry, based on ZOGY algorithm, yields results similar to our aperture photometry across all epochs.

This agreement between two independent reduction methods demonstrates that local background over-subtraction in aperture photometry is likely not the cause of the discrepancy with \citet{2026A&A...706A.154R} and \citet{2017ApJ...834..107B}. Instead, the offset in the bluest bands ($[3.6]/W1$) likely arises from subtle pipeline-level differences between difference-imaging implementations (e.g. PRF/PSF shape fitting versus aperture corrections, differences in epoch template frames, or residual background treatment on difference frames).
At $4.5~\mu\text{m}$ ($[4.5]/W2$), where host contamination is significantly dimmer and smoother, all methods converge to within $0.1$--$0.2$\,mag. Because the late-time SED of M101-OT is dominated by emission in the redder MIR bands ($[4.5]/W2$), the derived dust masses and temperatures in Section~\ref{subsec:sed_evolution} remain fully robust within reported uncertainties.

\section{NIR spectral line identification}

\begin{figure*}
    \centering
    \includegraphics[width=1.0\linewidth]{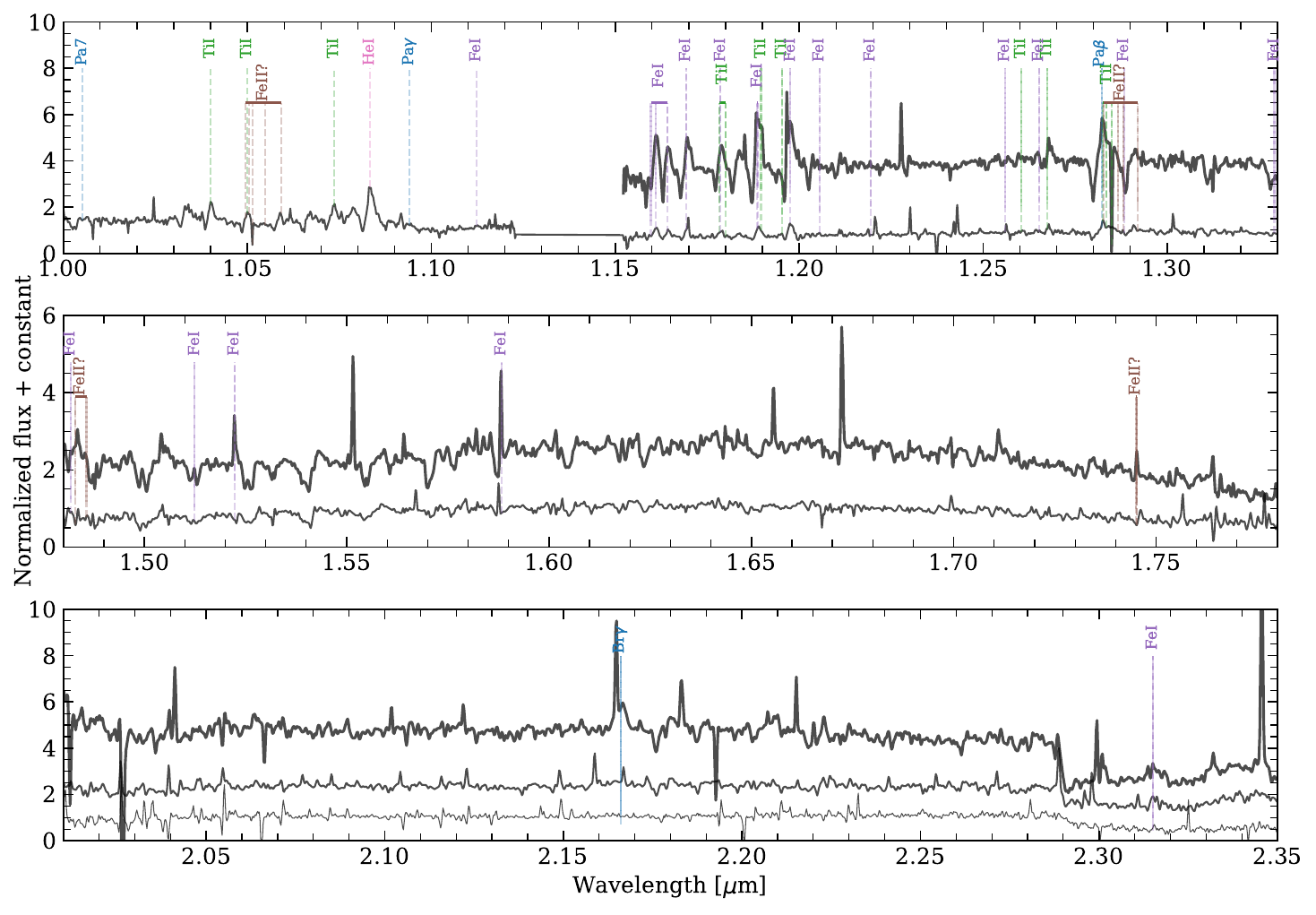}
    \caption{MOSFIRE spectra of M101-OT obtained in three different epochs at 31 March 2015 (+142\,d; thick solid line), 25 May 2015 (+197\,d; thin solid line), and 16 April 2016 (+524\,d; thinner line) covering the plateau phase after the second peak shown in $r$-band (see Fig.~\ref{fig:figure1}). The spectra were smoothed from the original resolution using a Gaussian kernel with a FWHM=4\,{$\AA$}. Identified atomic spectral features are indicated for \ion{Ti}{i}, \ion{Fe}{i}, possible \ion{Fe}{ii}, \ion{He}{i}, and \ion{H}{i} in the figure with different colours. The spectra are normalized, with the mean observed flux, and shifted for visualization purposes. \\
    (A colour version of this figure is available in the online journal.) \label{fig:C1}}
\end{figure*}

\end{appendix}

\clearpage

\end{document}